\documentclass[showkeys,showpacs,amsmath,superscriptaddress,preprintnumbers,amssymb,10pt,twocolumn]{revtex4-2}
\usepackage{graphicx}
\usepackage{dcolumn}
\usepackage{amsmath}
\usepackage{amssymb}
\usepackage{bm}
\usepackage{color}
\usepackage{subfigure}
\usepackage{ulem}

\begin{document}
\title{Effective spin dynamics of spin-orbit coupled matter-wave solitons in optical lattices}
\author{Kajal Krishna Dey}
\email{kajaldeypkc@gmail.com}
\affiliation{Department of Physics, Banwarilal Bhalotia College, Asansol-713303, West Bengal, India}
\author{Golam Ali Sekh}
\email{skgolamali@gmail.com}
\affiliation{Department of Physics, Kazi Nazrul University, Asansol-713340, West Bengal, India}
\begin{abstract}
We consider  matter-wave solitons in spin-orbit coupled Bose-Einstein condensates embedded in optical lattice and study  dynamics of soliton within the framework of Gross-Pitaevskii equations. We express  spin components of the soliton pair in terms of nonlinear Bloch equation and investigate effective spin  dynamics. It is seen that the effective magnetic field that appears in the Bloch equation is affected by the optical lattices, and thus  the optical lattice influences the precessional frequency of the spin components.  We  make use of numerical approaches to investigate the  dynamical behavior of density profiles and center-of-mass of the soliton pair in presence of the optical lattice. It is shown that the spin density is periodically varying due to flipping of spinors between the two states. The amplitude of spin flipping oscillation increases with lattice strength. We find that the system can also exhibit interesting nonlinear behavior for chosen values of parameters. We present a fixed point analysis  to study the effects of optical lattices on the nonlinear dynamics of the spin components. It is seen that the optical lattice  can  act as a control parameter  to change the dynamical behavior of the spin components from periodic to chaotic. 
\end{abstract}
\pacs{67.85.Hj, 03.75.Lm, 03.75.Kk, 67.85.Jk}
\keywords{Spin dynamics, Spin-orbit coupling, Optical lattice, Nonlinear Bloch equation, Regular and nonlinear dynamics, Chaos}
\maketitle
\section{Introduction}
The spin-orbit (SO) coupling plays an essential role in understanding several physical phenomena  which include quantum Hall effects\cite{hall1,hall2}, topological insulators\cite{hasan,kane} and topological superconductors\cite{scond,soc1}. It promises  applications in spintronics and topological quantum computing. In solid state material the presence of disorder and contaminant, however, make it difficult to observe such phenomena.  The remarkable experimental breakthroughs  on the realization of spin-orbit coupling (SOC) in Bose-Einstein condensate (BEC) of neutral atoms provides a platform for exploring SOC  in defect free medium with great flexibility\cite{zhan,thomas}. The  possibility to generate relatively stronger SOC in BEC leads to exotic phases due to non-conservation of spin part of the particles during the motion. {In addition, the BEC with angular spin-orbit coupling supports half-skyrmions in the ground state\cite{chen}. Breathing mode oscillation frequency of this skyrmion state is not universal but it depends on the strengths of SOC and nonlinear interaction\cite{balaz}.}

The inter-atomic interaction in Bose-Einstein condensates allows to explore nonlinear phenomena  such  as  rouge wave, dark soliton, gap soliton, bright soliton.  Under the action of Raman-induced SOC the dynamics of soliton changes appreciably.  The matter-wave soliton in a SOC-BEC suffers from the lack of Galilean invariance and induces  shape change  with the increase of velocity \cite{xu}. A SOC-BEC supports the so-called stripe solitons characterized by the density modulations in the form of stripes. Spatial motion of the soliton is coupled to the spin degree of freedom since the Raman transition can cause both momentum transfer and spin flipping.  It has recently  been demonstrated experimentally  that the center of mass  motion of a BEC in a harmonic trap is significantly affected by the spin degree of freedom\cite{morsch, giovanni}.
 
In this work, we comprehensively analyze the spin dynamics and center-of-mass motion of bright-bright solitons in spin-orbit coupled Bose-Einstein condensates in presence of optically induced periodic potential. We pay a special attention on the influence of spin degrees of freedom to the center-of-mass motion.  It is seen that the spin part of solution in presence of optical lattice can be represented on the Bloch sphere and the effective spin dynamics can be described by the so-called nonlinear Bloch equation\cite{wen,wen1}. Understandably,  an effective magnetic field is produced in the spinor  reference frame. This field causes the spinors to precess. The  precessional frequency is found to depend on  the lattice parameters and, hence, coupling between the spatial motion of soliton and its spin degrees of freedom enriches due to optical lattices. Specifically, the amplitude and frequency  of population oscillation due to spin flipping between the spin-orbit coupled states is modified in presence of OLs. 

The spin-orbit coupling  creates anharmonic characteristics in the collective spin dynamics in BECs with optical lattices(OLs). We employ a fixed point analysis both in presence and absence of OLs and find that the effective spin dynamics of the system is periodic  with a single period  in absence of optical lattices and the periodic multiplicity occurs in presence of optical lattices. This ultimately leads to chaos in the effective spin dynamics. {Chaotic dynamics of BECs in one-dimensional quasi-periodic optical lattice and random disordered potentials has been studied in \cite{swarup}.}  {Recently, it is demonstrated that the SOC may create anharmonic characteristics in collective dipole oscillations beyond the effective-mass approximation \cite{zhang,zhang1}.}

The paper is organized as follows. In section II, we introduce a set of coupled Gross-Pitaevskii equation to describe spin dynamics of two solitons in presence of spin-orbit interaction and optical lattices. Based on variational calculations we derive  equations for different  parameters  of solitons and hence find equations for the spin dynamics in presence of optical lattices using a similarity transformation.  We study the dynamics of effective spin components and center-of-mass from the variational calculation and also justify the variational predictions by the direct numerical simulation of the Gross-Pitaevskii equations in section III.  We envisage nonlinear spin dynamics of the soliton pair in section IV through the analysis of phase-space trajectories and Lypunov exponent in the parameter space\cite{chaos0,chaos1,chaos2}. Finally, we summarize the  results in section V.
\section{Theoretical formulation}
The spin-orbit interaction  generates coupling  between the two pseudo spin states created by using proper detuned Raman lasers in Bose-Einstein condensates. The two-component BECs thus created can be described by the coupled mean-field Gross-Pitaevskii equation(GPE). The  GPEs for spin-orbit coupled Bose-Einstein condensates in quasi-one-dimension is given by
\begin{eqnarray}
i\begin{pmatrix}
\Psi_t \\
\Phi_t
\end{pmatrix}
=\begin{pmatrix}
H_{11} & H_{12} \\
H_{21} & H_{22} 
\end{pmatrix}
\begin{pmatrix}
\Psi \\
\Phi
\end{pmatrix}.
\label{eq1}
\end{eqnarray}
Here
\begin{eqnarray}
H_{11}&=&-\frac{1}{2}{\partial_x^2}-i\beta\partial_ x +V(x)+c_{1}|\Psi|^2+c_{12} |\Phi|^2,\,\nonumber\\
H_{22}&=&-\frac{1}{2}{\partial_x^2}+i\beta\partial_ x +V(x)+c_{12} |\Psi|^2+c_{2}|\Phi|^2 \nonumber\\
{\rm and}\nonumber\\
H_{12}&=&H_{21}=\alpha 
\label{eq2}
\end{eqnarray}
with the effective potential
\begin{equation}
 V(x)=\frac{1}{2}\lambda_\perp^2 x^2+V_0 \cos(2 k_{\rm lat} x+\phi_{L}).
 \label{eq3}
\end{equation}
Here the first term stands for harmonic trap with $\lambda_\perp=\omega_x/\omega_\perp$, the ratio of longitudinal and transverse frequencies, and the second term gives optical lattice potential with amplitude $V_0$, wave number $k_{\rm lat}$  and phase $\phi_L$. In Eq. (\ref{eq2}), $\beta$ and $\alpha$ are the strengths of spin-orbit interaction and Raman coupling, strengths of inter and intra-component interactions are denoted by $c_j$ and $c_{ij}$ respectively.  We replace $t \rightarrow t/\omega_\perp$, $x \rightarrow x\,a_{\perp}$ and $V_0 \rightarrow V_0 \hbar \omega_\perp$ in Eq.(\ref{eq2}) such that it becomes dimensionless.

The SOC-BEC with attractive atomic interaction supports bright soliton solution in absence of optical lattices (OLs) and this soliton may contain nodes. In view of this, we consider the following trial solution of Eq.(\ref{eq1})\cite{wen}
\begin{eqnarray}
\begin{pmatrix}
\Phi\\
\Psi
\end{pmatrix}
=
\sqrt{a/2}
\begin{pmatrix}
\sin\theta\, {\rm sech}(a x+x_c)e^{i[p_{1x} x+\phi_{1x}]}\\
\cos\theta\, {\rm sech}(ax+x_c)e^{i[p_{2x} x+\phi_{2x}]}.
\end{pmatrix}
\label{eq4}
\end{eqnarray}
where $\theta$, $a$, $x_c$, $p_{jx}$ and $\phi_{jx}$ are variational parameters. Here $\theta$ and $a^{-1}$ determine respectively the population imbalance between two components and their width, $x_c$ is the center-of-mass, $p_{jx}$ and $\phi_{jx}$ represent wave number and  phase of the $j^{\rm th}$ component respectively. The trial solution  is  normalized to  $N=a/4$. We obtain  an averaged Lagrangian density  involving the variatonal parameters ($a$, $x_c$, $p_{1x}$, $p_{2x}$, $\phi_{1x}$ and $\phi_{2x}$) using  $\left<{\cal L}\right>=\int_{-\infty}^{+\infty} {\cal L} \,dx$.  It is given by
\begin{eqnarray}
\left\langle {\cal L}\right\rangle=\left\langle {\cal L}_1\right\rangle+\left\langle {\cal L}_2\right\rangle+\left\langle {\cal L}_3\right\rangle+\left\langle {\cal L}_4\right\rangle
\end{eqnarray} 
where
\begin{subequations}
\begin{eqnarray}
\left\langle {\cal L}_1\right\rangle&=&-\pi V_0\frac{\cos(2 k_{lat}x_c/a+\phi_L)}{\sinh(\pi/a)},
\end{eqnarray}
\begin{eqnarray}
\left\langle {\cal L}_2\right\rangle&=&-\frac{\pi \alpha}{2}\cos\left[\frac{x_c}{a}p_{x}+\phi \right]\frac{p_{x}\sin 2\theta}{\sinh(\pi p_{x}/2a)}\nonumber\\
&+&\beta_1\left({p}_{1x}\sin^2\theta-{p}_{2x}\cos^2\theta\right),
\end{eqnarray}  
\begin{eqnarray}
\left\langle {\cal L}_3\right\rangle&=&\left(\frac{1}{2}\,p^2_{2x}-\frac{x_c}{a}\dot{p}_{2x}+a\dot{\phi}_{2x}\right)\cos^2\theta \nonumber\\ &-&\left(\frac{1}{2}p^2_{1x}
-\frac{x_c}{a}\dot{p}_{1x}+a\dot{\phi}_{1x}\right)\sin^2\theta,
\end{eqnarray}
\begin{eqnarray}
\left\langle {\cal L}_4\right\rangle\!\!&=&\!\!\!-\frac{a^2}{6}(a\!+\!c_1 \cos^4\!\theta\!+\!c_2\sin^4\!\theta\!+\!\frac{c_{12}}{2} \sin^2 \!2\theta),
\end{eqnarray}
\end{subequations}
where $k_n=(p_{1x}-p_{2x})/2$, $k_p=(p_{1x}+p_{2x})/2$  and $\phi_n=\phi_{1x}-\phi_{2x}$. We make use of the Ritz optimization procedure and obtain the following equations for the variational parameters (see Appendix). 
\begin{subequations}
\begin{eqnarray}
\frac{d \left\langle z\right\rangle}{dt}&=&k_p,
\label{eq7a}
\end{eqnarray}
\begin{eqnarray}
\dot{k}_p&=&\!\frac{2\pi \alpha \lambda^2 \sin \phi}{a\sinh(\pi \lambda/a)} \sin 2\theta +\frac{2\pi V_0\lambda^2 \sin \phi}{a \sinh(\pi \lambda/a)},
\end{eqnarray}
\begin{eqnarray}
\dot{\phi}&=&\!2\lambda k_p\! -\!\frac{2\pi \alpha \lambda \cos\phi}{a \sinh(\pi \lambda/a)} \cot 2\theta,
\end{eqnarray}
\begin{eqnarray}
\dot{\theta}&=&-\frac{\pi \alpha \lambda}{a \sinh(\pi \lambda /a)}\sin\phi.
\label{eq7d}
\end{eqnarray}
\end{subequations}
Here $\phi=2k_n<z>+2\phi_n$, $<z>=-x_c/a$, $\beta\approx k_n$, $k_n=\lambda$, $k_{lat}\approx k_n$, $\phi_L\approx 2\phi_n$, $c_1=c_2=c_{12}$ and $\dot{k}_n=0$. Equations (\ref{eq7a})-(\ref{eq7d}) describe the dynamics of different parameters of soliton solutions of the system. Particularly, Eq. (\ref{eq7d}) shows how the population imbalance between the soliton components varies with the phase difference. In this context we note that  this equations is also useful to study Josephson-type oscillation in SOC-BECs\cite{sumaita, bloch}.

\subsection{Effective equations for the spin components}
Let us consider normalized complex-valued spinors: $\Psi_j=\sqrt{\rho(x,t)}\chi_j$ with $\chi=(\chi_\uparrow,\chi_\downarrow)$ and  $\Psi_j=(\Psi,\Phi)$ such that $\rho=|\Psi|^2+|\Phi|^2$ and $|\chi_{\uparrow}|^2+|\chi_{\downarrow}|^2=1$, where $\chi_\uparrow=\sin\theta\,e^{i\phi_{1x}}$ and $\chi_\downarrow=\cos\theta\,e^{i\phi_{2x}}$. Spin expectation value can be defined through the transformation
\begin{eqnarray}
{\bf S}=\chi^T{\sigma}\chi.
\end{eqnarray}
Here $\sigma\equiv \{\sigma_x,\sigma_y,\sigma_z\}$ is the set of Pauli spin operators. Expectation values of different spin components are given by
\begin{subequations}
\begin{eqnarray}
S_x&=&\chi^T\sigma_x\chi=
\chi^*_\uparrow \chi_\downarrow+\chi^*_\downarrow \chi_\uparrow=\sin 2\theta \cos \phi\\
S_y&=&\chi^T\sigma_y\chi=\chi^*_\downarrow \chi_\uparrow-\chi^*_\uparrow \chi_\downarrow=-\sin 2\theta \sin\phi\\
S_z&=&\chi^T\sigma_x\chi=|\chi_\uparrow|^2-|\chi_\downarrow|^2=-\cos 2\theta.
\end{eqnarray}
\end{subequations}
Equations for the dynamics of spin components are obtained from Eq.(9) as
\begin{subequations}
\begin{eqnarray}
\dot{S}_z=2\tilde{\Omega} \,S_y,
\label{eq10a}
\end{eqnarray}
\begin{eqnarray}
\dot{S}_y=-2\tilde{\Omega} S_z-2\lambda k_p S_x,
\label{eq10b}
\end{eqnarray}
\begin{eqnarray}
\dot{S}_x=2 \lambda k_p S_y,
\label{eq10c}
\end{eqnarray}
\end{subequations}
where $\tilde{\Omega}=\frac{\pi\alpha\lambda}{a \sinh(\pi \lambda /a)}$ and 
\begin{eqnarray}
{k}_p=-\lambda S_z- \frac{2 \lambda V_0}{\alpha} \sin^{-1}(\tilde{S}_z)  +c_0
\label{eq11}
\end{eqnarray}
with $\tilde{S}_z=\sqrt{(1+S_z)/2}$ and  $c_0=\lambda S_{z0}+\frac{2 \lambda V_0}{\alpha}\sin^{-1}(\tilde{S}_{z0})$. Here we take $\left<\dot z(0)\right>=k_p(0)=0$. Note that $\bm{a}=(S_x,S_y,S_z)$ represents a vector on the Bloch sphere  and it satisfies,  $\bm{\rho}=(1+ \bm{a}\cdot\bm{\sigma})$. 

The Eqs. (\ref{eq10a})-(\ref{eq10c}) for the spin dynamics  can be expressed in the form
\begin{eqnarray}
 \dot{\bm{ S}}={\bm{ S}}\times {\bm{ B}}
 \label{eq12}
\end{eqnarray}
with ${\bm{B}}=\left(-2\tilde{\Omega},0,2\lambda k_p\right)$. This is the so-called nonlinear Bloch equation. Thus we see that  spinors $\chi_\uparrow$ and $\chi_\downarrow$ face an effective  magnetic field ${\bm{ B}}$\cite{eugen}. The field causes spin precession and the frequency of which depends on the lattice  and SOC parameters. Understandably, the frequency of precession is $\omega_p=\gamma |{\bm{B}}|=(4\lambda^2 k_p^2+4\tilde{\Omega}^2)^{1/2}$, where  the effective gyromagnetic ratio $\gamma=1$. Therefore, the spin parts of the solutions can be expressed as $\chi_\uparrow=\sin\theta\, e^{i(\phi_{1x}+\omega_p t)}$ and $\chi_\downarrow=\cos\theta\, e^{i(\phi_{2x}-\omega_p t)}$.

In order to study the dynamics of the spin component, we combine  (\ref{eq10a})-(\ref{eq10c}) and (\ref{eq11}), and write
\begin{eqnarray}
\ddot{S}_z + m_1 S_z+m_2 S_z^2 +m_3 S_z^3+m_4=V_L({S}_z)
\end{eqnarray}
with
\begin{equation}
V_L({S}_z)= \sin^{-1}(\tilde{S}_z)\left[v_1\!+v_2 {S}_z +\! v_3 S_z \!\sin^{-1}(\tilde{S}_z)\right]. 
\label{eq15}
\end{equation}
Here  $m_1=-2 c_1 \tilde{\Omega}\lambda^2+4 \tilde{\Omega}^2+2\lambda^2 c_0^2$, $m_2=4 \lambda^3 c_0$, $m_3=2 \lambda^4$ and $m_4=2\lambda\tilde{\Omega}c_0 c_1$ with $c_1=S_{x0}-S_{z0} (\lambda k_p/\tilde{\Omega})$. In Eq. (\ref{eq15}), $v_1=2c_1V_0 \lambda^2\tilde{\Omega}/\alpha$, $v_2=-8 V_0 \lambda^3(c_0+\lambda)/\alpha$ and $v_3=-8 V_0^2 \lambda^4/\alpha^2$. We see that for $V_0=0$,  the frequency of oscillation of $S_z$ under harmonic approximation is $\sqrt{m_1}$. In presence of the optical lattices,  the oscillation frequency further changes to $\sqrt{m_1+v_1}$ in the linear limit. In the nonlinear limit, one can, however, expect richer spin dynamics due to anharmonic response of the system.  In the following we consider both regular and nonlinear spin-dynamics in presence of optical lattices.

\section{Regular spin dynamics of BB-type solitons in optical lattices}
\begin{figure}[h!]
\includegraphics[width=4.2cm, height=3cm]{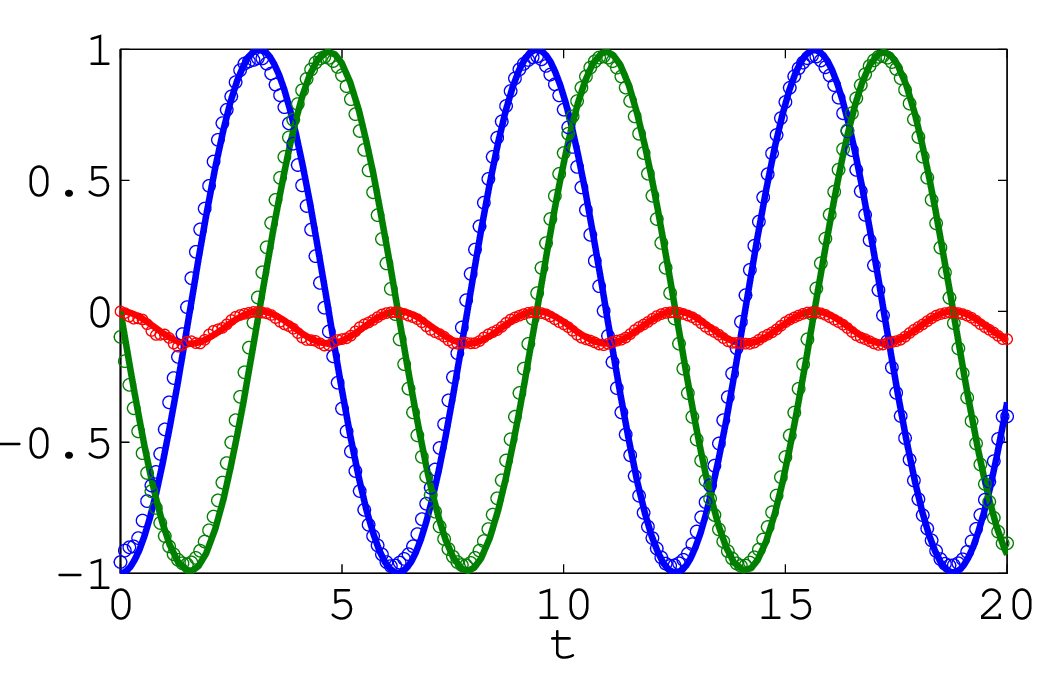}
\includegraphics[width=4.2cm, height=3cm]{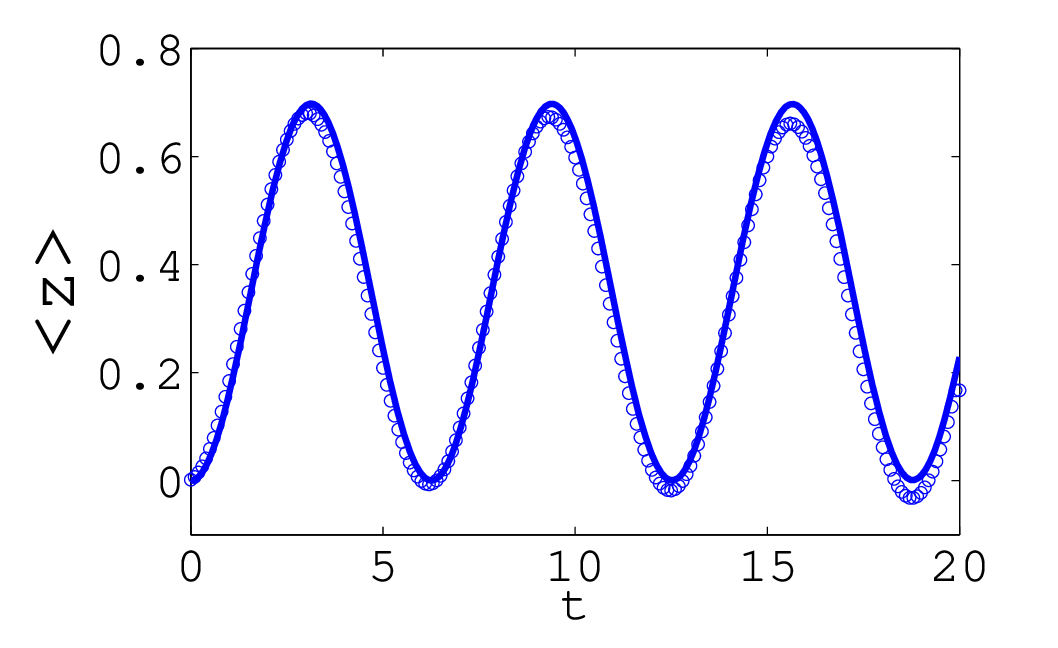}
\includegraphics[width=4.2cm, height=3cm]{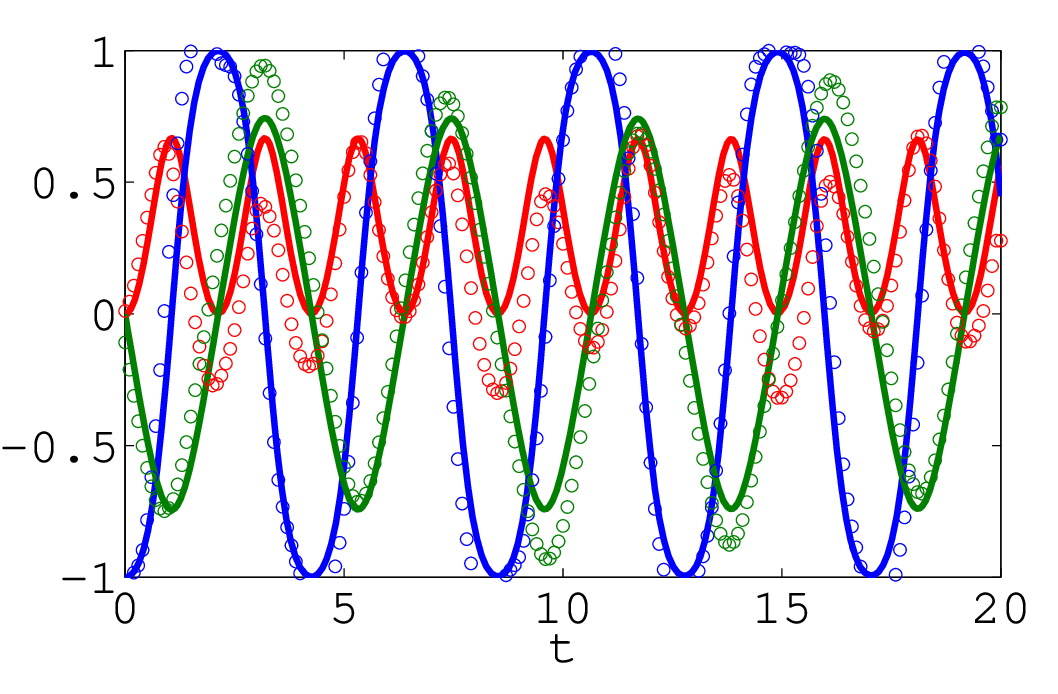}
\includegraphics[width=4.2cm, height=3cm]{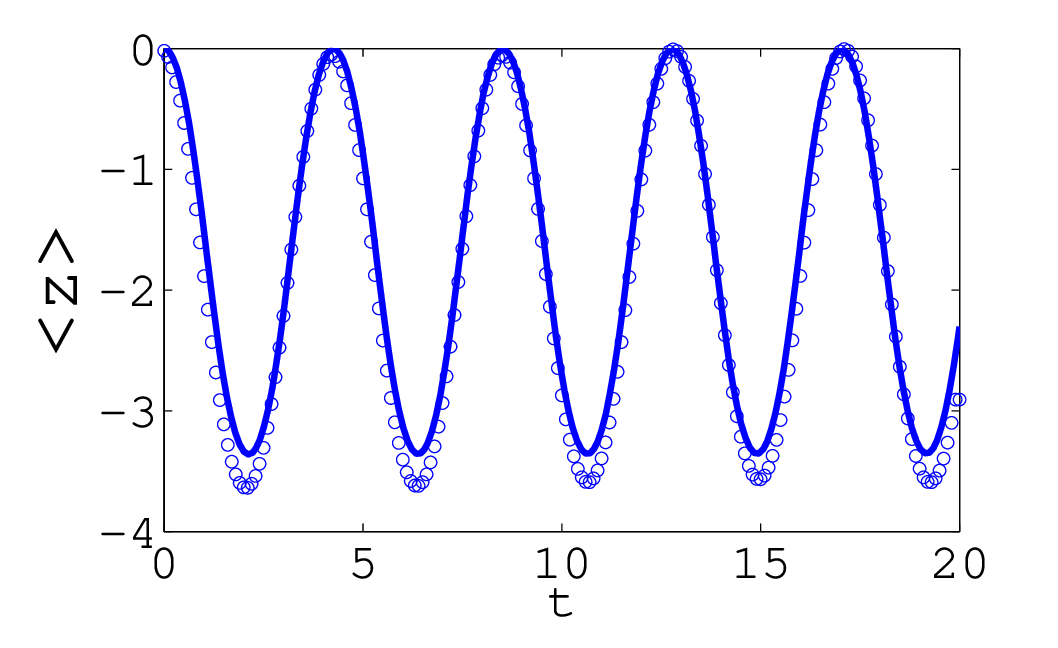}
\caption{Top panel:  Spin components of the solitons for the initial conditions $\theta(0)=\pi/4$ and $\phi(t=0)=\pi/4$.  In the left figure, the red line represents $S_{x}$,  the blue line represents $S_{y}$ and the green line represents $S_{z}$.  In the right panel gives  the evolution of centre-of-mass coordinate of the solitons. Here the initial state is chosen  for $\alpha=0.5$, $\beta=0.5\sqrt{\alpha}$, $c=-10$, $V_{0}=0$. Bottom panel: Same as those shown in the top panel but in presence of OLs with $V_0=-5$.
In both the panels, solid lines are obtained from variational calculations and circles are generated by direct numerical simulations of GPE.}
\label{fig1}
\end{figure}

Note that the centre-of-mass dynamics of the solitons pair is governed by Eq.(\ref{eq7a}) where the parameter $k_p$ is a function of spin expectation value. From Eq.(\ref{eq11}) and Eq.(\ref{eq7a}), we find an effective equation
\begin{eqnarray}
\frac{d^2 \left\langle z\right\rangle}{dt^2}=-2\lambda \tilde{\Omega} S_y+ \frac{2 \lambda \tilde{\Omega} V_0}{\alpha} \sin(\phi).
\label{eq15a}
\end{eqnarray}
Clearly, Eq.(\ref{eq15a}) describes a dependence of center-of-mass motion on the expectation value of spin components. It is seen that the center-of-mass is accelerated  due to  the force $-2\lambda \tilde{\Omega} S_y$ provided by the spin-orbit coupling. The solitons  encounter also a linear restoring force due to the optical lattice potential. 
To illustrate the dynamics of spin expectation value and center-of-mass in detail, we  numerically solve the  Eqs.(\ref{eq10a})-(\ref{eq10c}) and  Eq. (\ref{eq7a}) for $\theta(t=0)=\pi/4$ and $\phi(t=0)=\pi/4$. The results obtained for the spin dynamics are shown  in left panel of  Fig.\ref{fig1} while in right panel  we show the time evolution of  centre-of-mass.   Here, we fix initial values of all the spin components of system on the  Bloch sphere such that $\sum_{j=x,y,z} S_j^2=1$. The figure clearly shows that expectation values of the spin components oscillate   periodically with time due to SOC,. However,  the periodicity and amplitudes of oscillations of different spin-components are not same.  We see that the centre-of-mass of the system  also oscillates periodically. 

\begin{center}
\begin{figure}[h!]
\includegraphics[width=3cm, height=2.5cm]{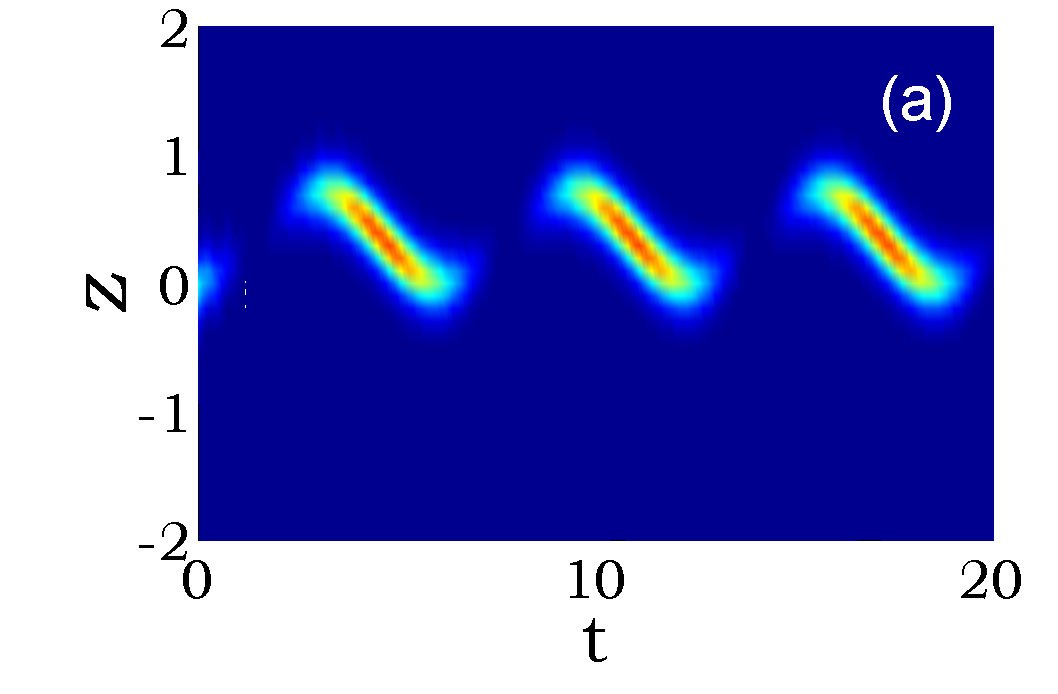}
\hspace{-0.5cm}
\includegraphics[width=3cm, height=2.5cm]{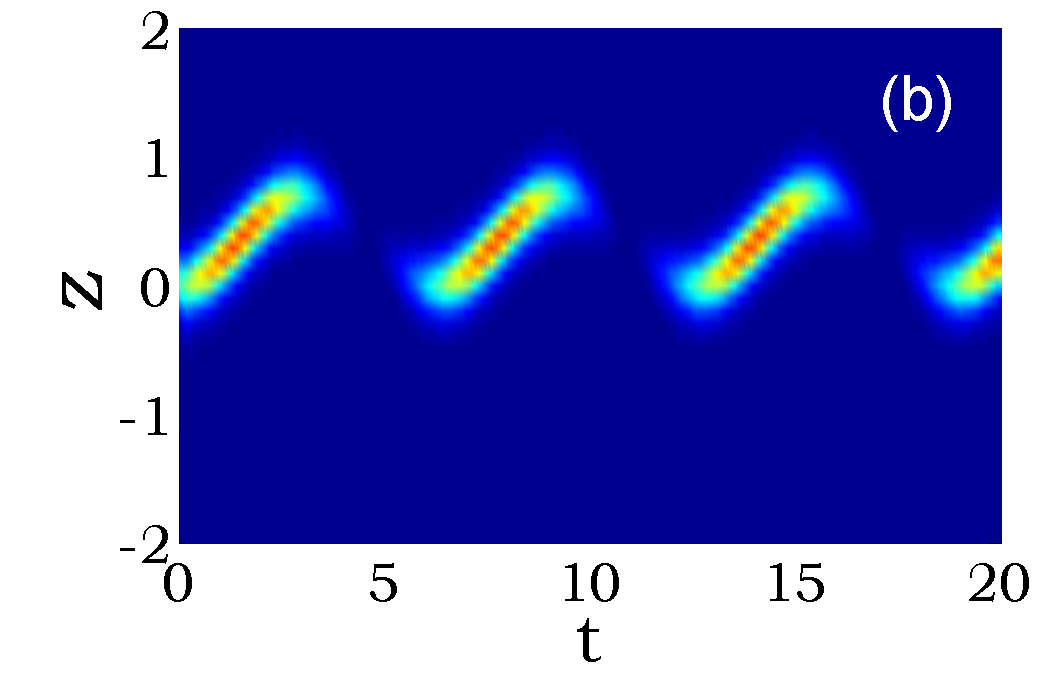}
\hspace{-0.5cm}
\includegraphics[width=3.0cm, height=2.5cm]{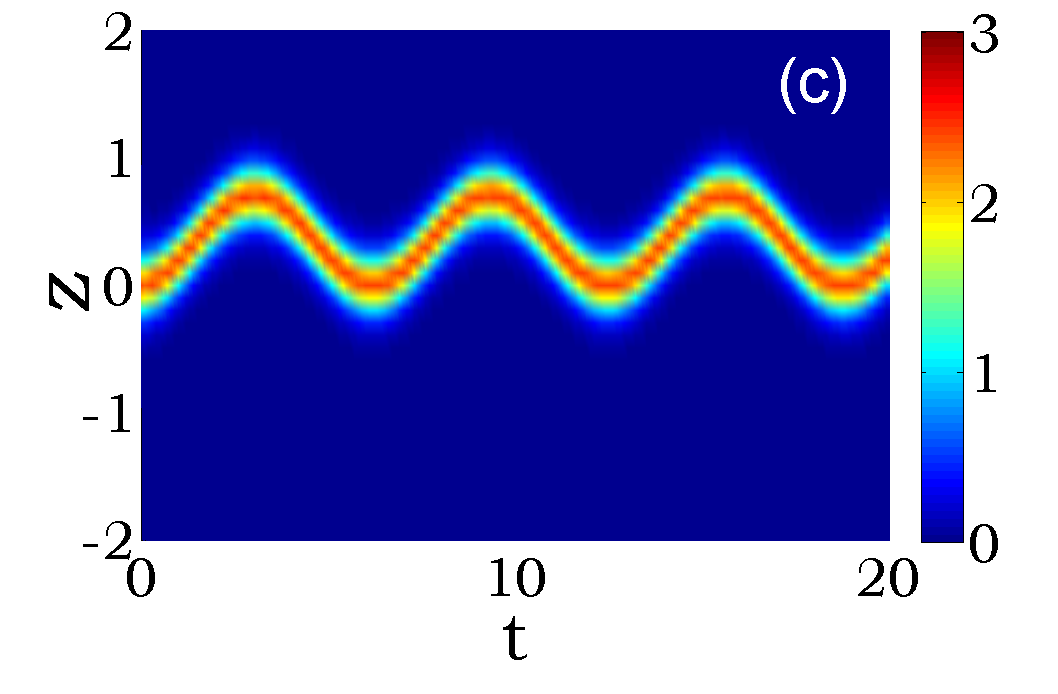}
\includegraphics[width=3cm, height=2.5cm]{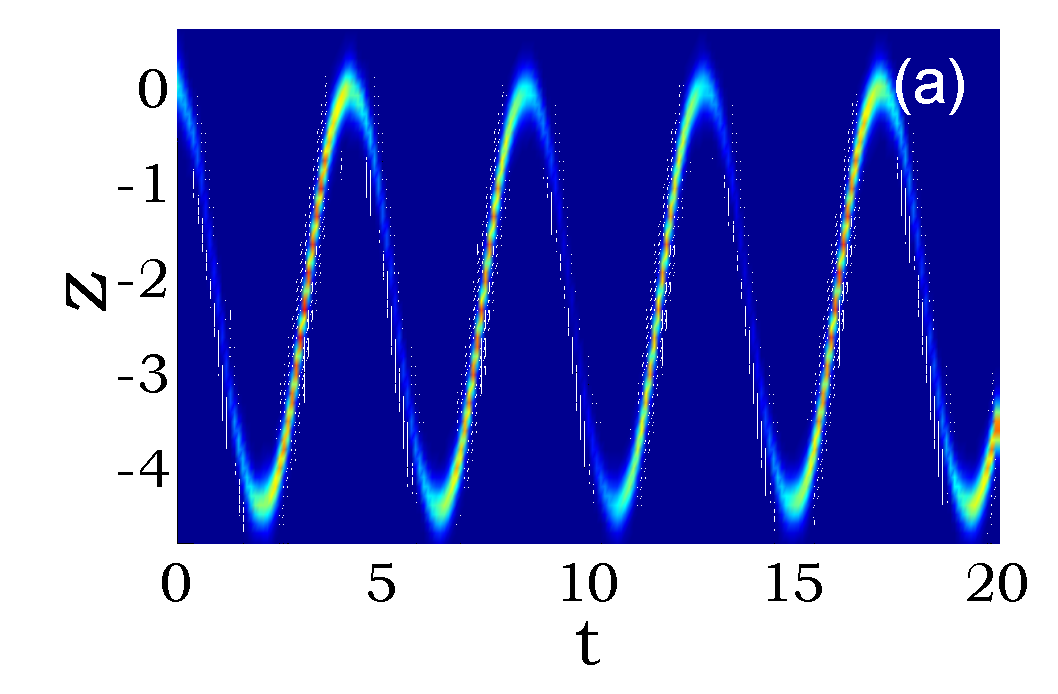}
\hspace{-0.5cm}
\includegraphics[width=3cm, height=2.5cm]{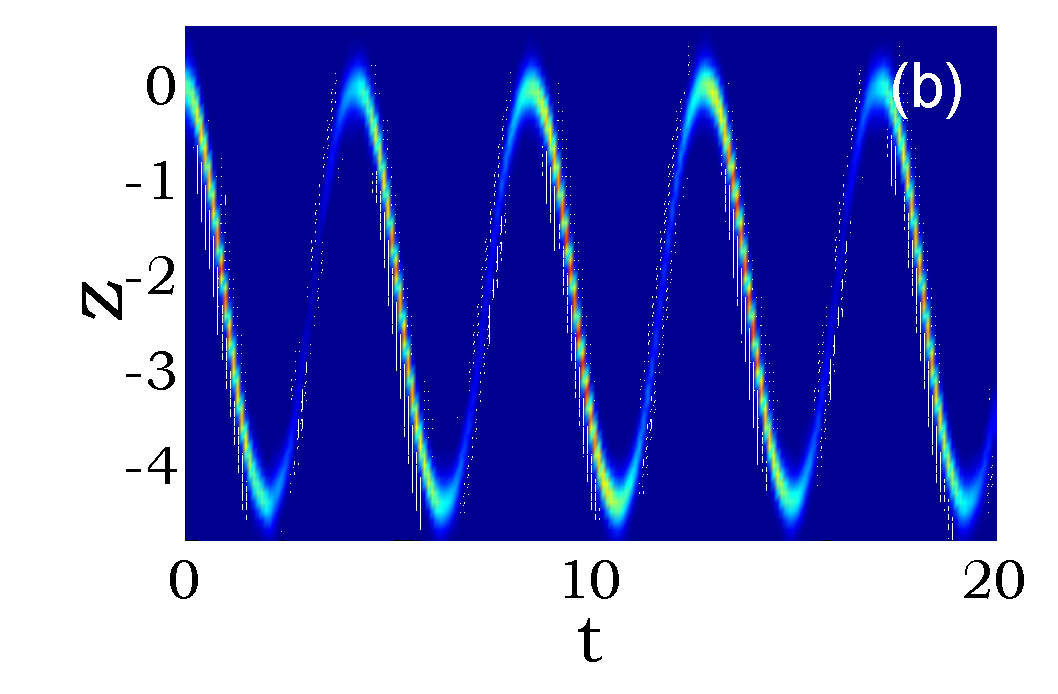}
\hspace{-0.5cm}
\includegraphics[width=3.0cm, height=2.5cm]{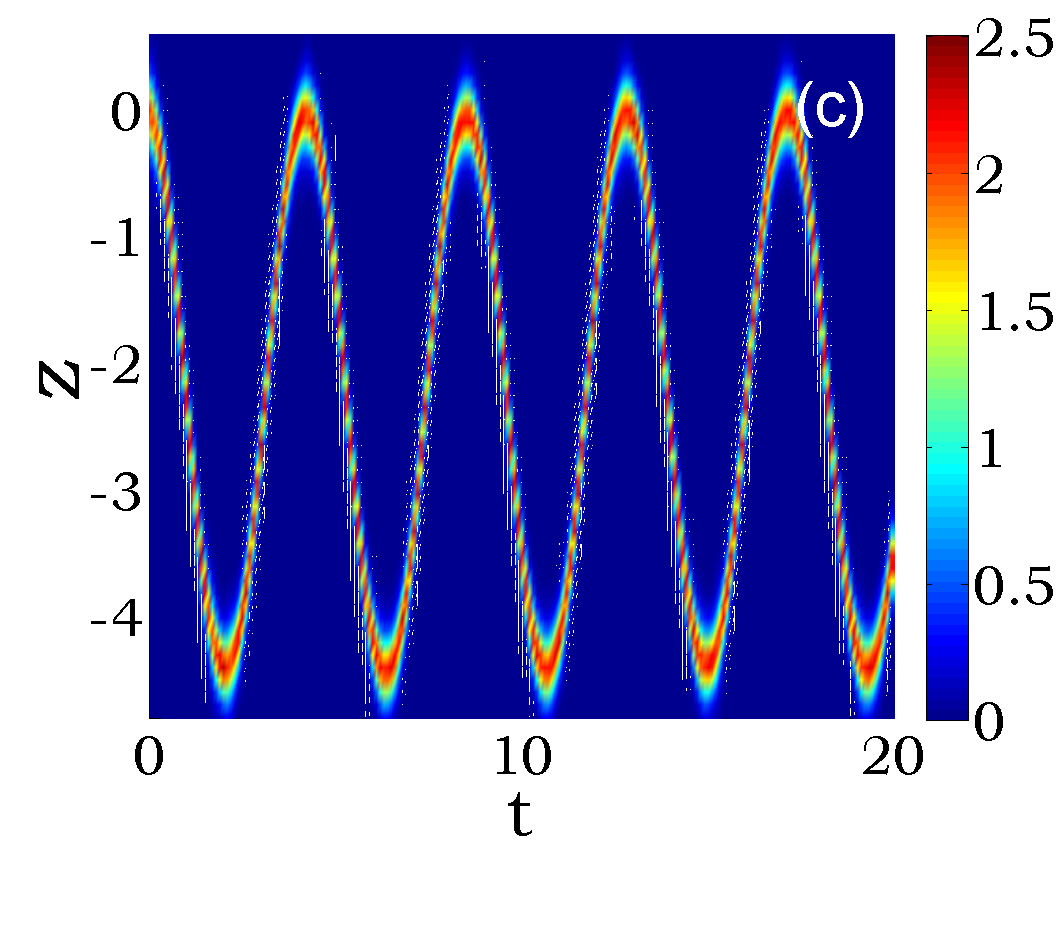}
\caption{Top panel: Evolution of densities  of both components produced by the GPE simulations with the initial conditions $\theta(t=0)=\pi/4$, $\phi(t=0)=\pi/4$, $\alpha=0.5$, $\beta=0.5\sqrt{\alpha}$, $c=-10$ and $V_{0}=0$. Here the panel marked by (a) gives density ($|\psi|^{2}$) of the first component while the panel (b) represents the density($|\phi|^{2}$) of the second component. The total density($|\psi|^{2}+|\phi|^{2}$) variation is shown in the panel (c). 
Bottom panel: The same as that shown in the top panel  but $V_{0}=-5$.}
\label{fig2}
\end{figure}
\end{center}

It is seen that the centre-of-mass of solitons oscillates due to SO coupling and the expectation values of spin components move periodically in a closed orbit on the Bloch sphere. Since  the spin dynamics can influence the center-of-mass motion and the later is  expected to be affected by the optical lattices,  the oscillations spin component are also expected to be affected by the external periodic potential. In order to find the  effect of optical lattice on the spin dynamics and hence also on the centre-of-mass coordinate of the solitons, we take $V_0=-5$. In the bottom panel of Fig. \ref{fig1}, we show the variations of expectation value of spin components (left panel) and of centre-of-mass (right panel) for the interaction strength $c=-10$. It is seen that the amplitude and frequency of center-of-mass motion increases with the introduction of optical lattice. The optical lattice also causes phase shift in the center-of-mass dynamics. The periodic dynamics of spin components are affected by lattice potential. Particularly, the amplitude of oscillation of only $x$-component of spin changes significantly due to optical lattice. 

{In order to authenticate the results obtained from the variational approximation, we numerically solve the Gross-Pitaevskii equation(GPE) in Eq.(\ref{eq1}). Time-dependent GPE is discretized in space and time using the split-step Crank-Nicolson scheme. Real-time propagation is employed to solve the discretized equation to study the BEC dynamics by using sufficiently small space and time steps \cite{ravisankar, muruganandam}.} Taking the ansatz of BB-type solitons in Eq.(\ref{eq4}) as initial condition($t=0$) for $p_{1x}=-p_{2x}=\lambda$  and $x_c=0$ we calculate time evolution of $S_{x}$, $S_y$, $S_z$ and $<z>$ both in absence and presence of optical lattices(OLs).

A comparison of results obtained from variational and direct numerical simulations is presented in Fig.\ref{fig1} where the circles are generated by direct GPE numerical simulations. We see that the direct numerical simulations of GPE agree very well with those obtained from variational calculations. Both the studies infer that both the dynamics of spin expectation values and center-of-mass of soliton pair are periodic in time. The numerical simulation of GPE with SOC in presence of OLs also show good agreement  with the variational results. More specifically, modifications of amplitudes, phase and frequency of spin components and center-of-mass in OLs are clearly reflected in both the cases.

We calculate spatial variation of density distributions with time of both the components from the direct numerical simulation of the GPE. It is seen  that density varies periodically with time. The numerical simulation confirms that the density variation occurs due to periodic interchange of particles between the components (Fig.\ref{fig2}). As a result, the value of spatial densities of two components alternately disappear and revive keeping the total density constant. Interestingly, the variation of total density with time in space is also periodic keeping the density constant. We also note that, for a comparatively weaker atomic interaction, the solitons decay under the action of spin-orbit coupling.

In presence of OLs, the density of the solitonary wave oscillates in space with relatively larger frequency. During the motion densities of  both the components  vary with time and the variations are not independent of each other. Particularly, if the density is large in the first half period in one component then it is large in the second half period in second component. In this way the density of two components alternately decays and revives with time. However, the total spatial density remains constant. This implies that the  spin flip-flop takes place periodically and causes variation to the population density of each component. The density varies largely with a larger frequency as the  strength of lattice increases. {The striking feature is that the variation of spatial density shows an additional collective periodic variation. From Eq.(\ref{eq15}) it is evident that the center-of-mass directly depends on the optical lattice parameter. Therefore, an appropriate strength of lattice can also result such additional collective variation in the density of each component.} 

\section{Nonlinear spin dynamics of BB-type solitons in optical lattices}
The precession of soliton spin is described by the nonlinear Bloch equations, in which the nonlinear terms mostly result from SO coupling. We can see that the spin precession couples to the centre-of-mass motion under the action of SO coupling. A stronger lattice potential results in an increase of effective magnetic field in the $z$-direction and thus augments the precession frequency. Optical lattice potential acts as a small perturbation to this nonlinearity which may have a significant impact on the nonlinear spin dynamics. 

\begin{figure}[h!]
	\includegraphics[width=4.2cm, height=3.5cm]{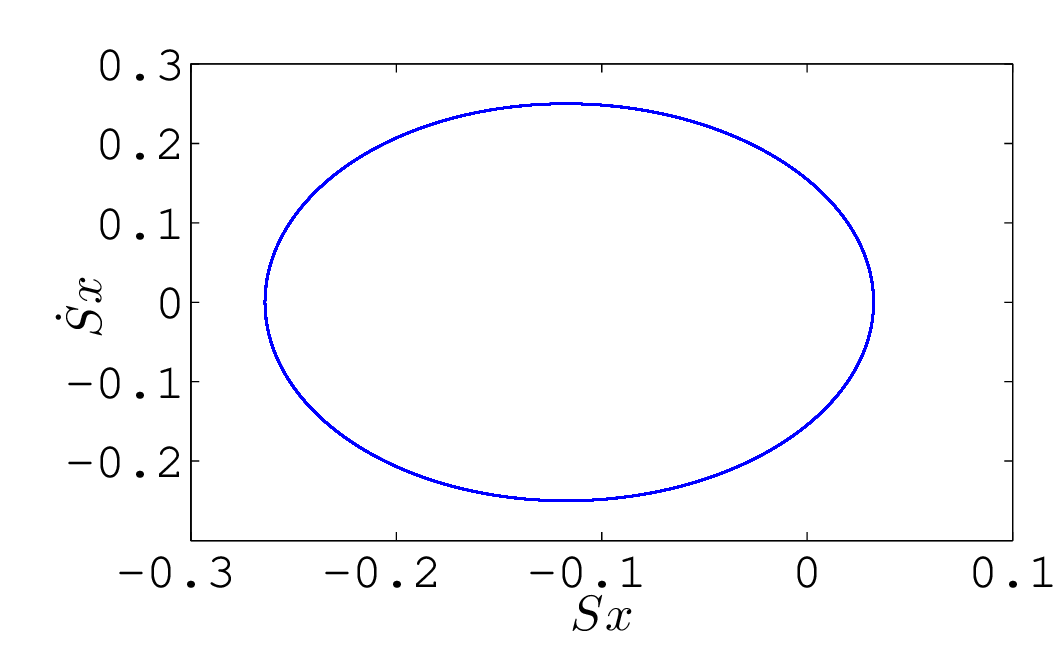}
	\includegraphics[width=4.2cm, height=3.5cm]{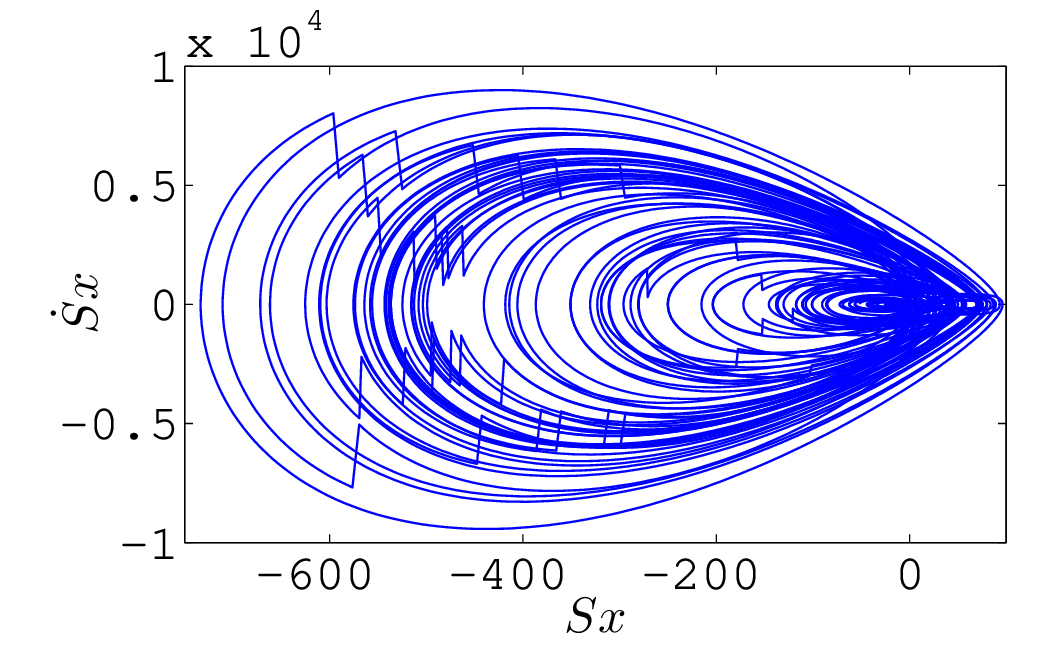}
	\hspace{0.5cm}
 	\includegraphics[width=4.2cm, height=3.5cm]{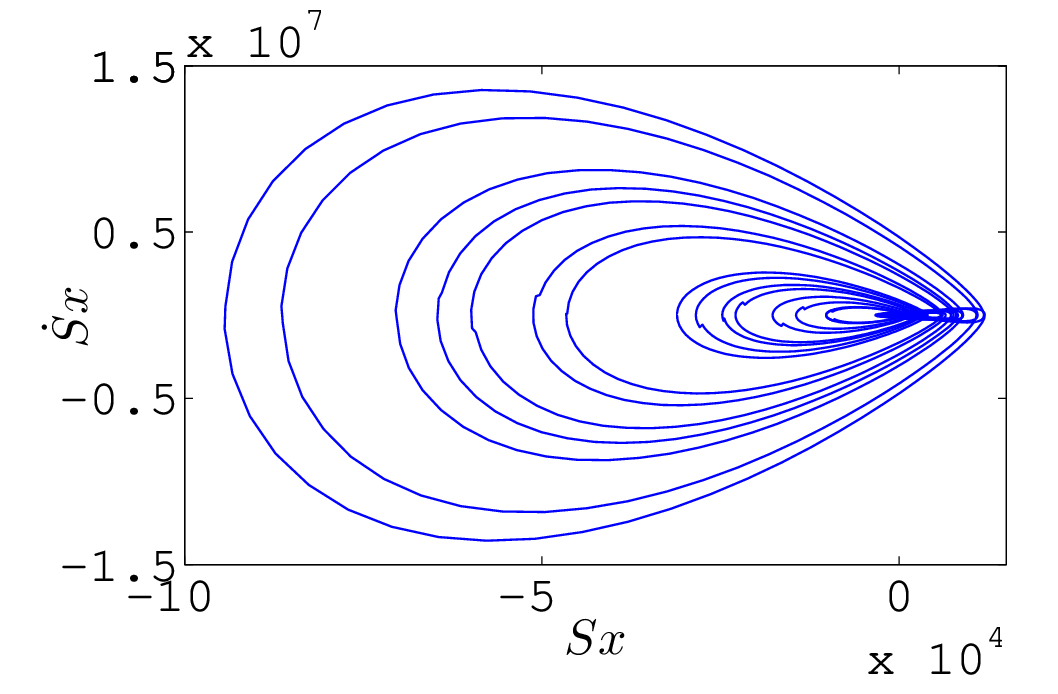}
	\includegraphics[width=4.2cm, height=3.5cm]{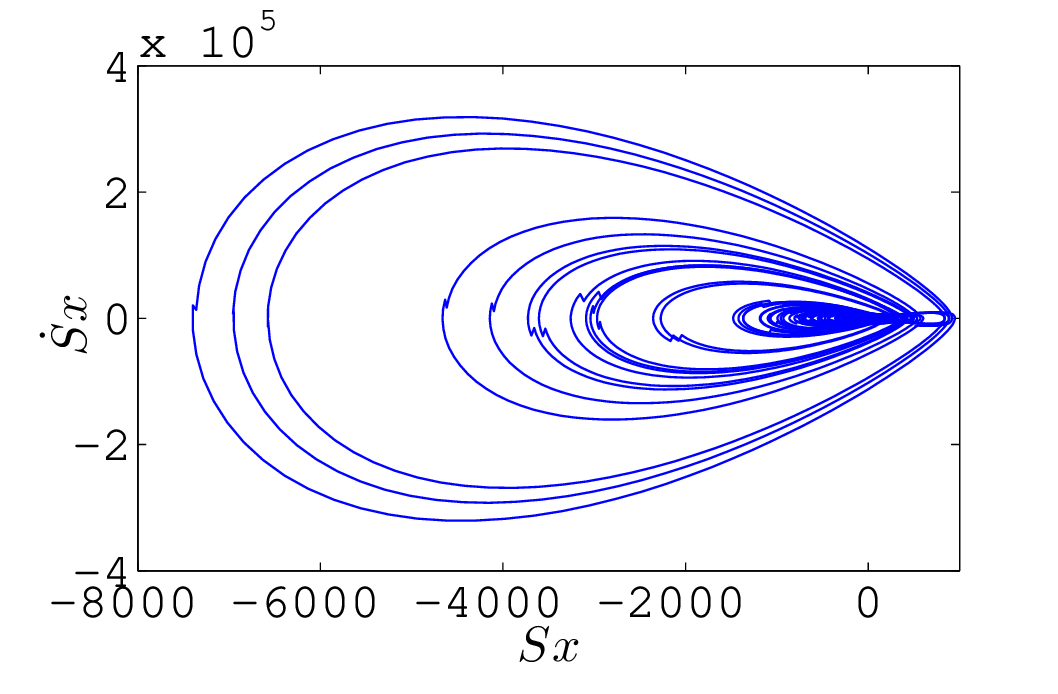}
	\caption{Structure of phase space trajectory of spin component $S_{x}$ for (i) $c=-2, V_{0}=0$(top left panel), (ii) $c=-2, V_{0}=-1$ (top right panel) (iii) $c=-2, V_{0}=-2$(bottom left panel) and (iv) $c=-4, V_{0}=-1$(bottom right panel).}
	\label{fig4}
\end{figure}

\begin{figure}[h!]
	\includegraphics[width=4.2cm, height=3.5cm]{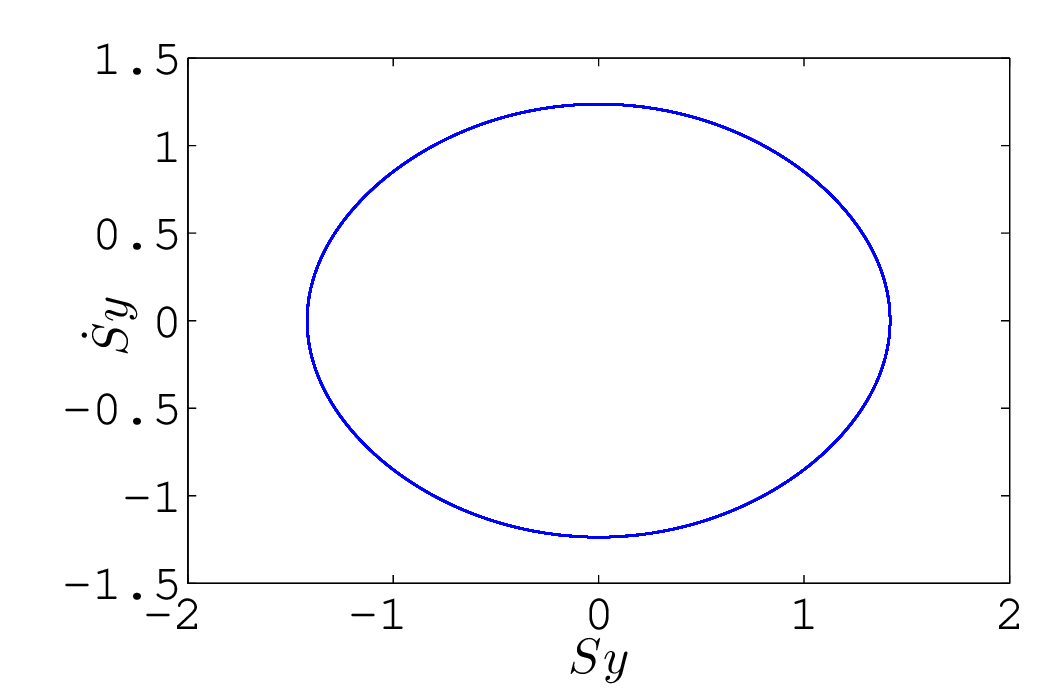}
	\includegraphics[width=4.2cm, height=3.5cm]{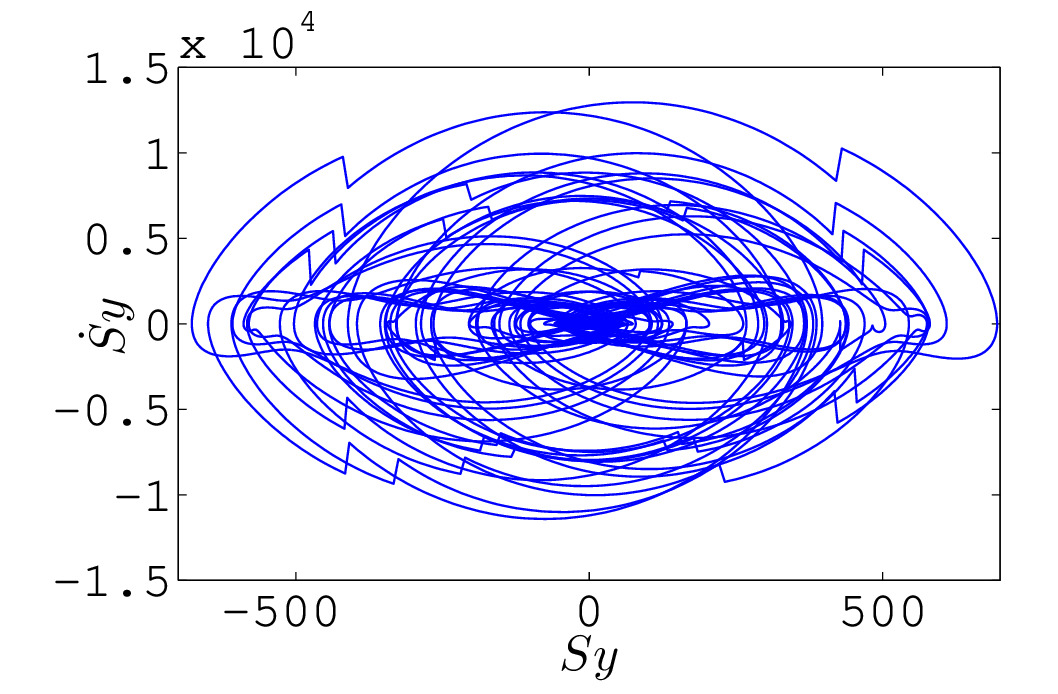}
	\includegraphics[width=4.2cm, height=3.5cm]{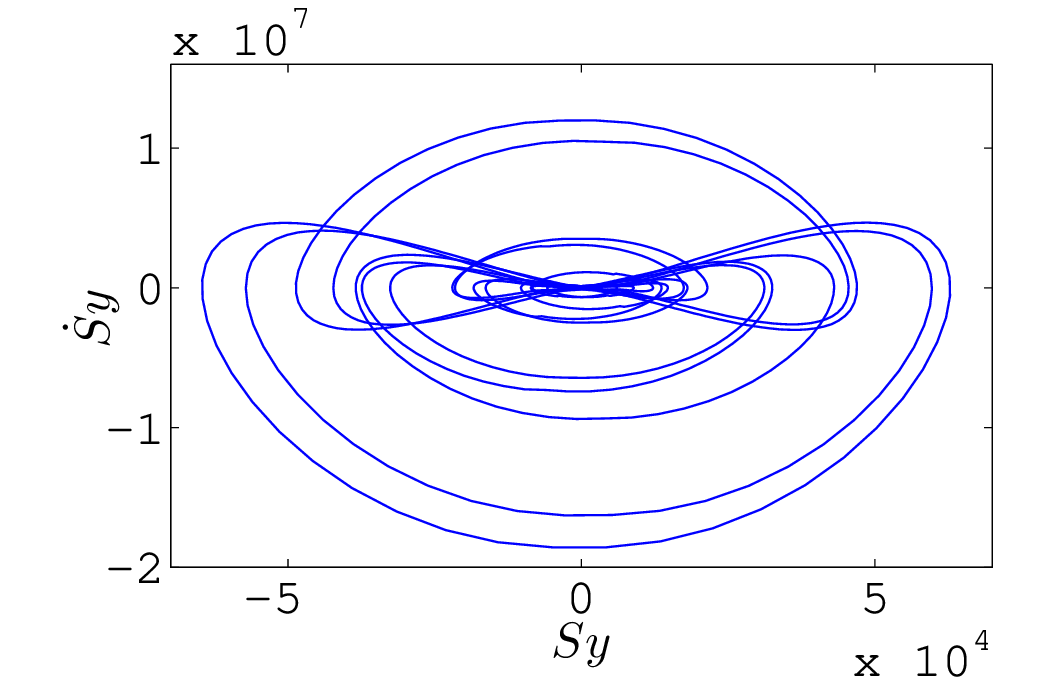}
	\includegraphics[width=4.2cm, height=3.5cm]{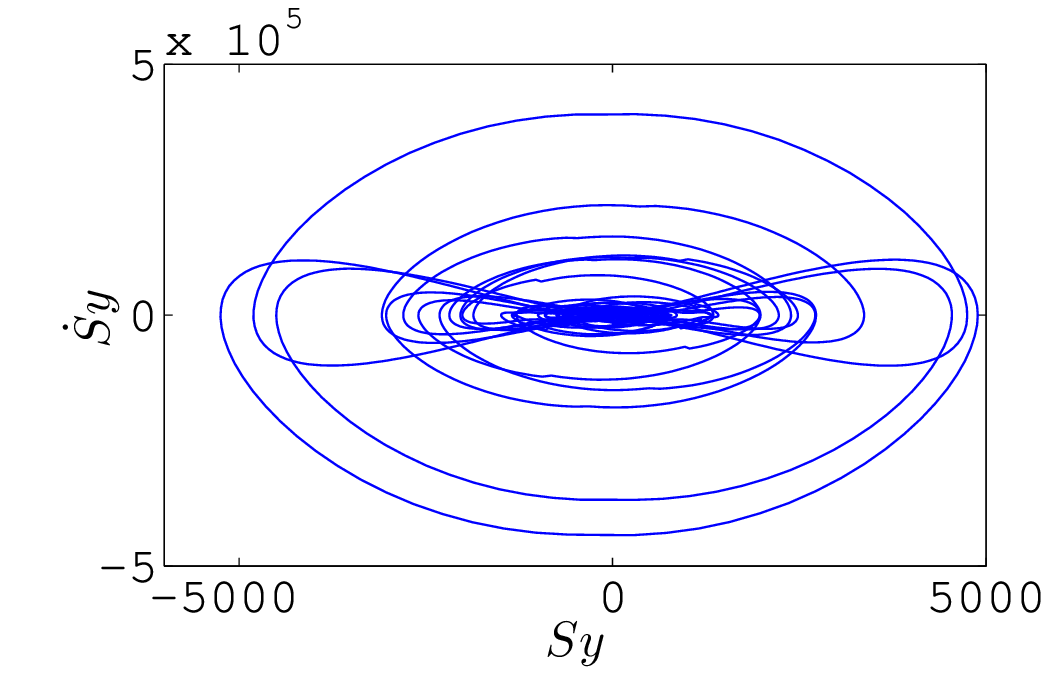}
	\caption{Structure of phase space of spin component $S_{y}$ for (i) $c=-2, V_{0}=0$(top left panel), (ii) $c=-2, V_{0}=-1$(top right panel), (iii) $c=-2, V_{0}=-2$(bottom left panel) and (iv) $c=-4, V_{0}=-1$(bottom right panel).}
	\label{fig5}
\end{figure}

\begin{figure}[h!]
	\includegraphics[width=4.2cm, height=3.5cm]{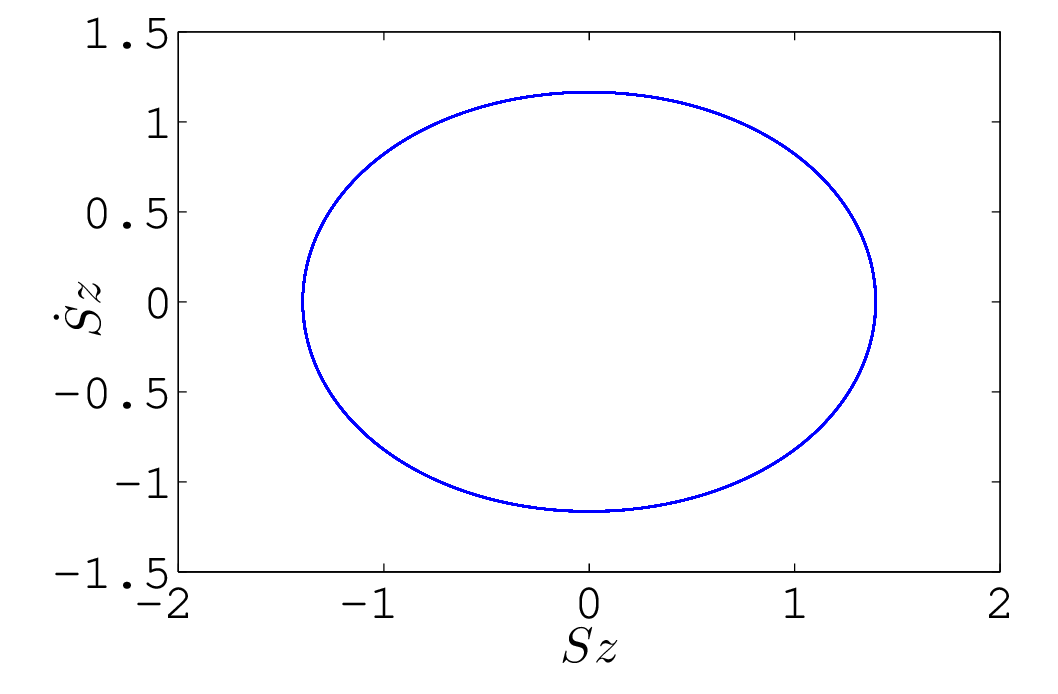}
	\includegraphics[width=4.2cm, height=3.5cm]{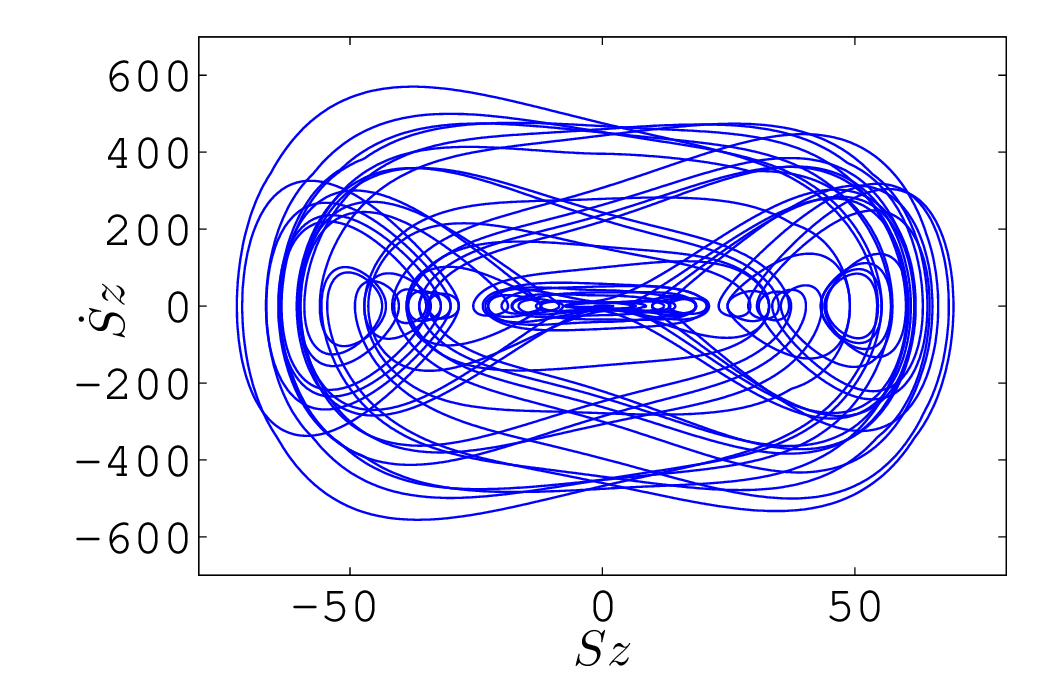}
	\includegraphics[width=4.2cm, height=3.5cm]{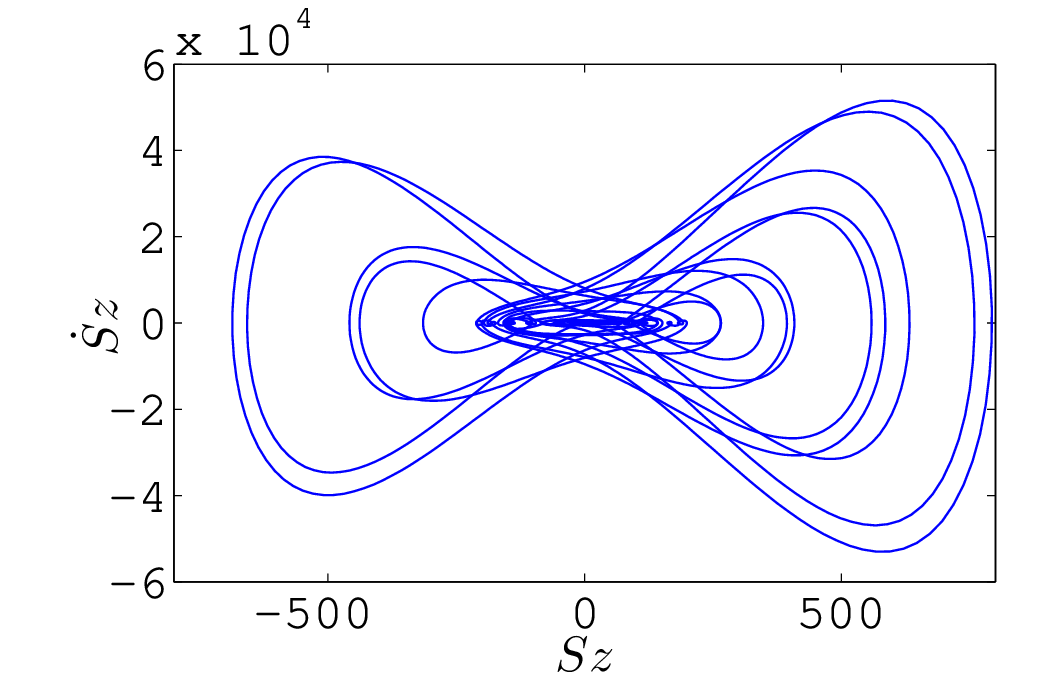}
	\includegraphics[width=4.2cm, height=3.5cm]{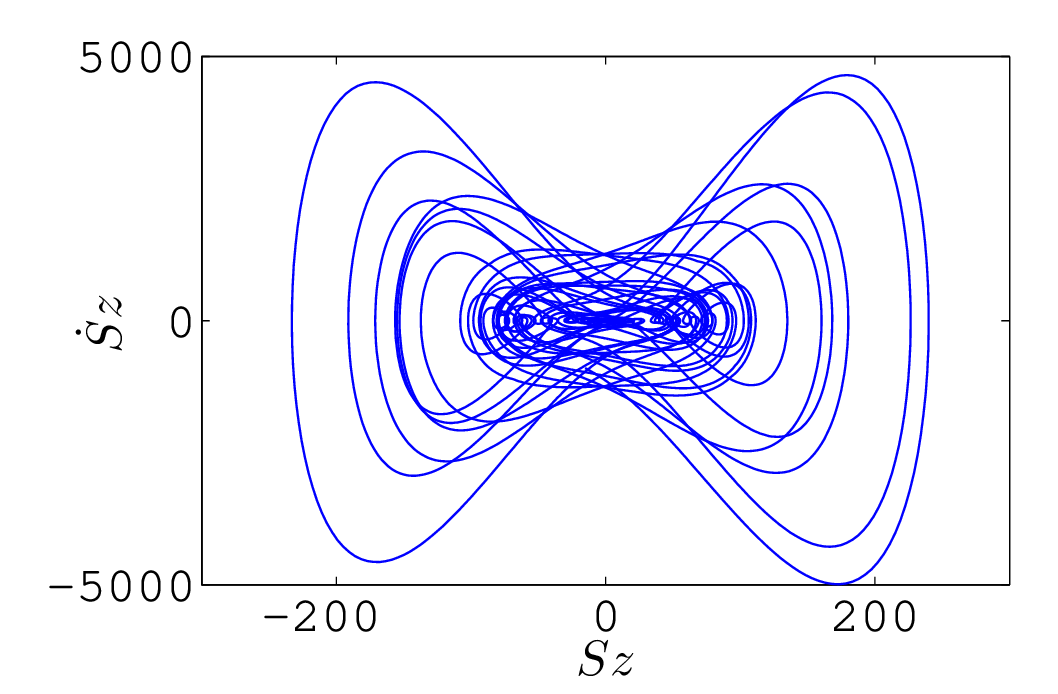}
	\caption{Structure of phase-space of spin component $S_{z}$ for (i) $c=-2, V_{0}=0$(top left panel), (ii) $c=-2, V_{0}=-1$( top right panel), (iii) $c=-2, V_{0}=-2$(bottom left panel) and  (iv) $c=-4, V_{0}=-1$(bottom right panel).}
	\label{fig6}
\end{figure}

In order to study the influence of the lattice potential on the nonlinear spin dynamics of BB-type solitons, we solve the coupled nonlinear Bloch  equation(\ref{eq12}) numerically taking initial condition at the fixed point ${\bm S}_c=(S_{\rm xc},S_{\rm yc},S_{\rm zc})$ of the system. The fixed point is determined by solving the Bloch equation at steady state\cite{shs}. More specifically, we make a fixed point analysis for the spin components in absence ($V_0=0$) and presence ($V_0\neq 0$) of optical lattices and display the outcome in Figs.\ref{fig4}-\ref{fig6}. The phase plot of $x$, $y$ and $z$ components of the effective spin ${\bm S}$ for zero lattice strength are periodic with period $1$. The period of phase-space trajectory increases with the increase of lattice strength. It  roughly resembles with the famous R\"ossler system where periodic doubling occurs with the increase of a control parameter\cite{rossler}. In the present case, the lattice strength may be considered as a control parameter. We see that the optical lattice introduces periodic multiplicity and it is increases with the increase of lattice strength. This ultimately results in chaos in the dynamics of spin-components. We also notice that a prominent chaotic dynamics can be observed for a weaker lattice strength if inter-atomic interaction is taken relatively stronger. {Contrarily, the trajectory of chaotic pattern diverges dramatically with the rise of lattice potential while maintaining the same interaction strength.} 

\begin{figure}[h!]
	\hspace{-0.5cm}
	\includegraphics[width=4.2cm, height=3.5cm]{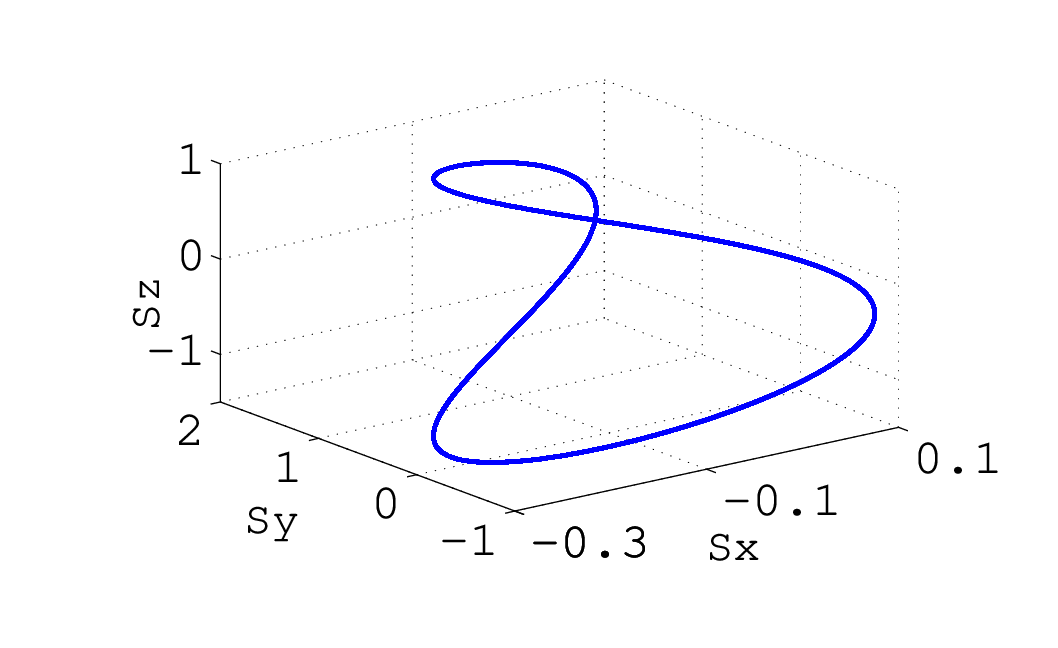}
	\hspace{-0.5cm}
	\vspace{-0.5cm}
	\includegraphics[width=4.2cm, height=3.5cm]{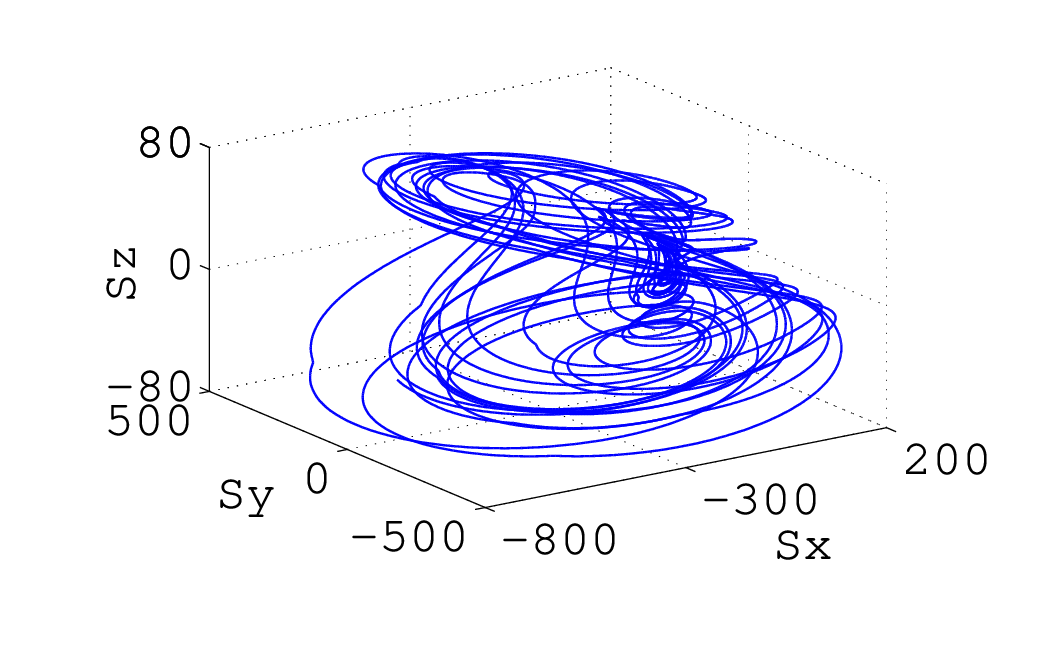}
	\includegraphics[width=4.2cm, height=3.5cm]{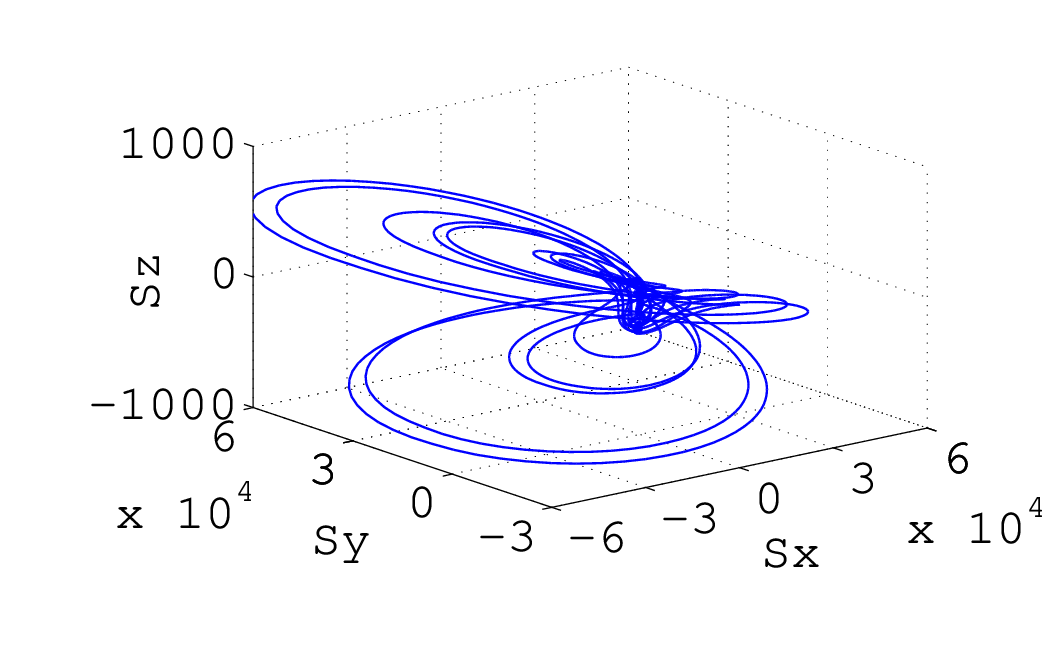}
	\includegraphics[width=4.2cm, height=3.5cm]{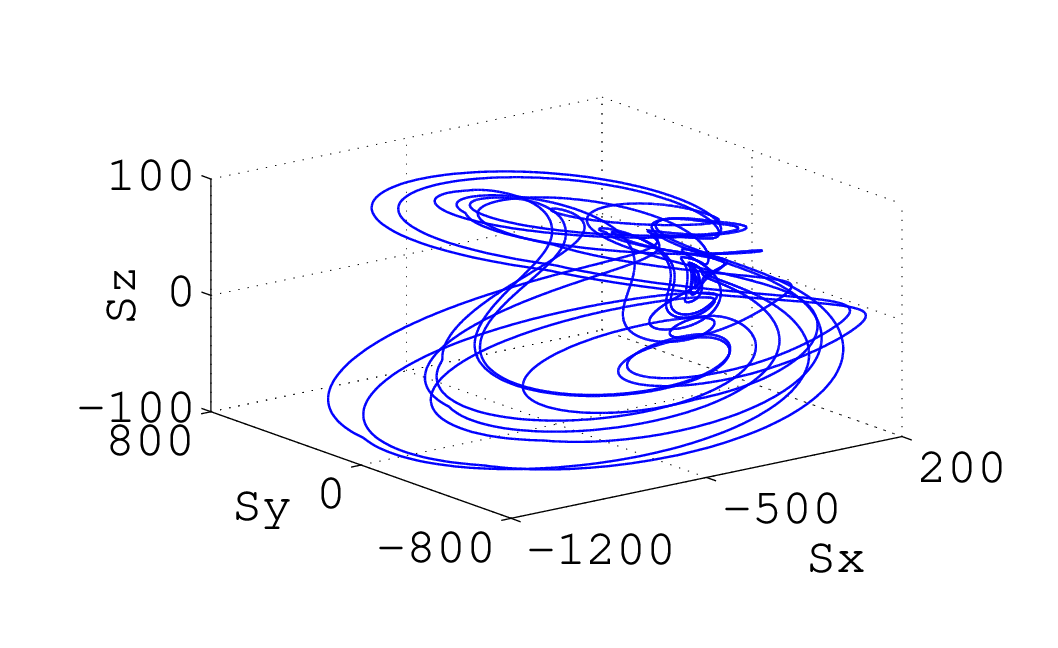}
\caption{Phase space trajectory of three spin components with the initial conditions $S_{x}=-0.12,  S_{y}=-1,  S_{z}=-1$. for (i) $c=-2, V_{0}=0$(top  left panel), (ii)  $c=-2, V_{0}=-1$(top right panel), (iii) $c=-2, V_{0}=-2$ (bottom left panel) and (iv) $c=-4, V_{0}=-1$(bottom right panel).}
	\label{fig6a}
\end{figure}

The variation of  spin expectation values, $S_x,S_y$ and $S_z$  in phase-space for different values of lattice and interaction parameters is displayed in Fig.\ref{fig6a}. We see that the relative change of spin values in  phase-space  is double periodic in absence of OLs ($V_0=0$) and the periodic multiplicity increases with the increase of lattice strength($V_0\neq 0$). For a relatively stronger value of nonlinear interaction, the time evolution of spin projections in phase-space also becomes multi-period. This gives  an indication  to chaotic dynamics of spin components.

\begin{figure}[h!]
	\includegraphics[width=4.2cm, height=3.5cm]{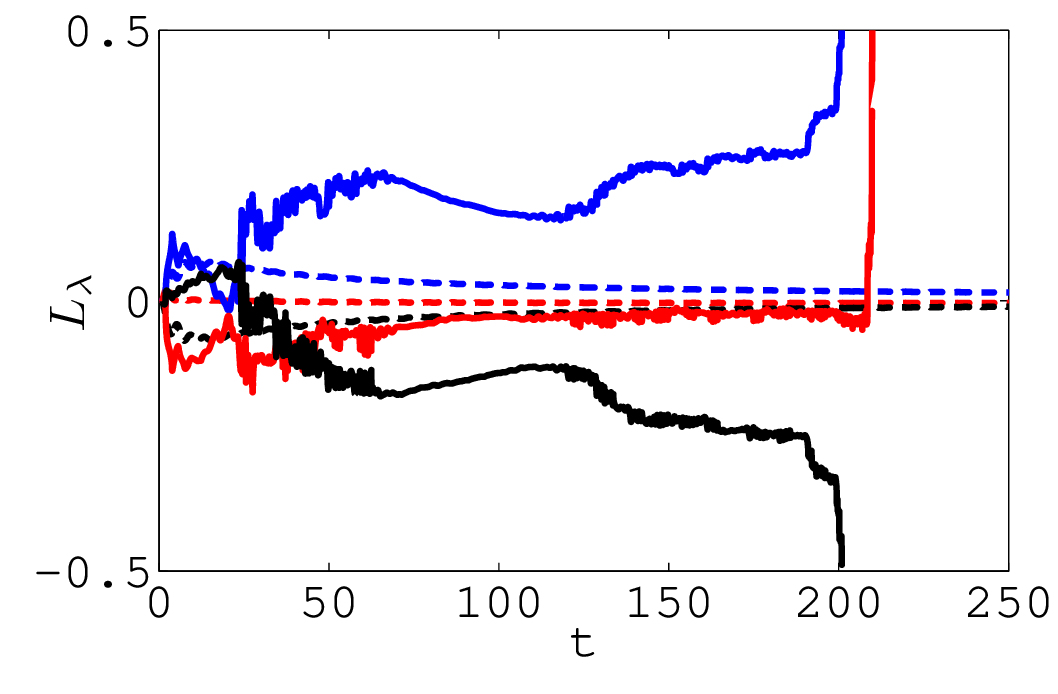}
	\includegraphics[width=4.2cm, height=3.5cm]{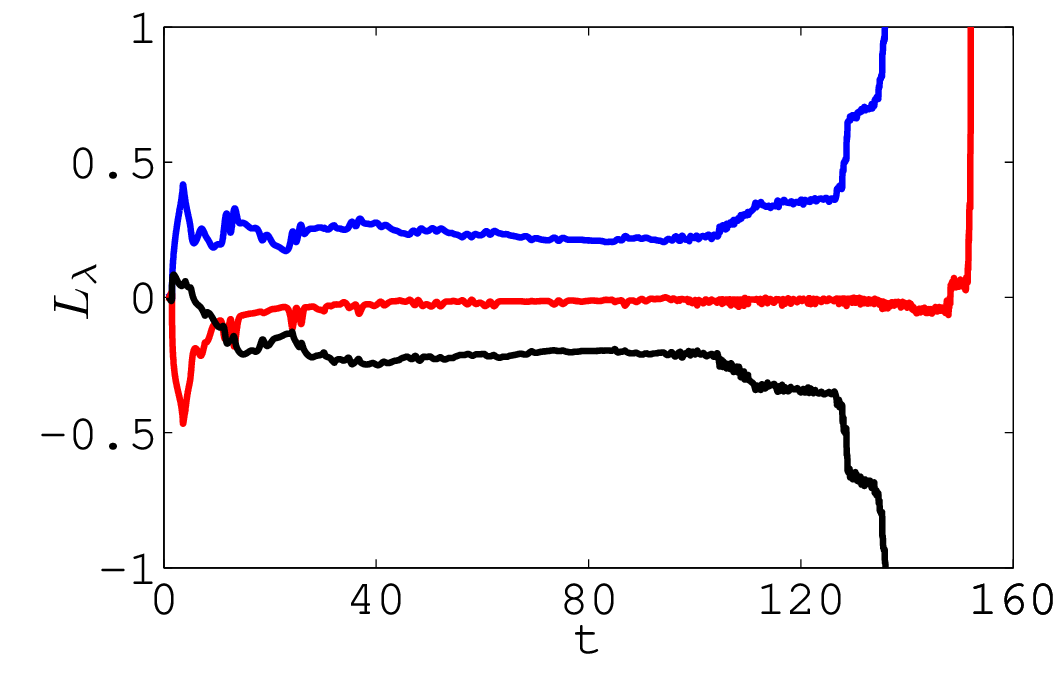}
	\includegraphics[width=4.2cm, height=3.5cm]{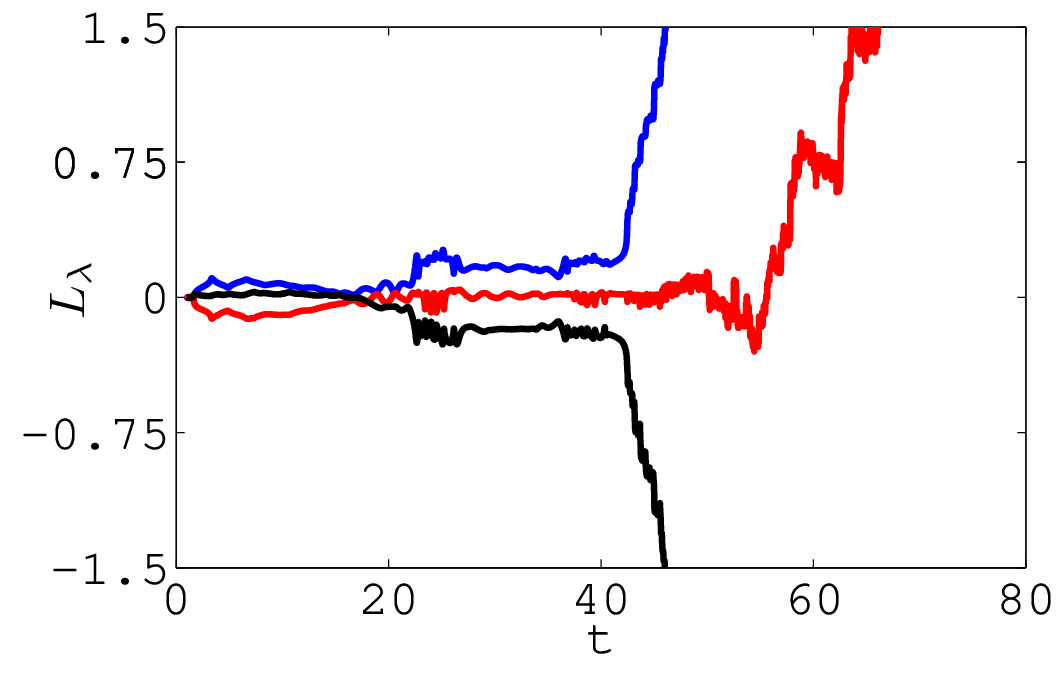}
	\includegraphics[width=4.2cm, height=3.5cm]{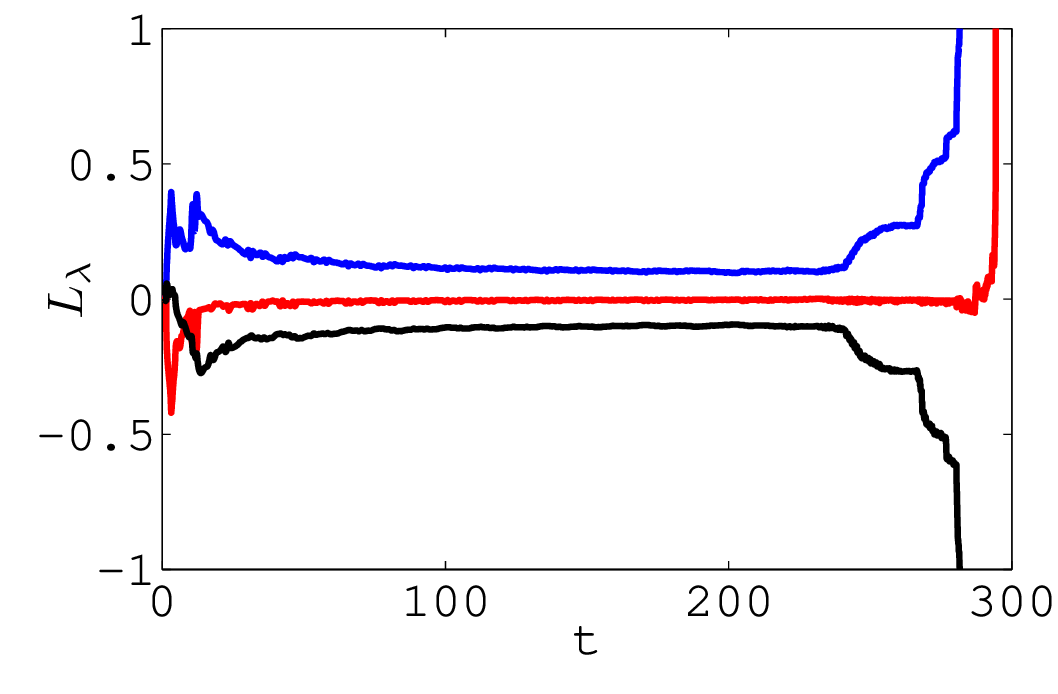}
	\caption{Lyapunov exponent $(L_\lambda)$ of spin components. Blue curve represents $S_{x}$ while red and black curves give $S_{y}$ and $S_{z}$ respectively.  Top left panel: It gives the result for(i) $c=-2, V_{0}=-1$(solid curves) (ii)  $c=-2, V_{0}=0$(dashed curves). Top right panel: It gives the results for $c=-2, V_{0}=-2$. Bottom left panel: It shows $L_\lambda$ for $c=-4, V_{0}=-1$. Bottom right panel: It gives $(L_\lambda)$ for $c=-4, V_{0}=-2$.}
\label{fig7} 
\end{figure}

In order to check our prediction from the phase-space pattern of spin dynamics, we calculate the Lyapunov exponent(LE) at different times both in the presence and absence of lattice potential for the given values of interaction strengths. The results are shown in Fig. \ref{fig7}.  It shows that  the LEs for all the spin components  remain close to zero for $V_0=0$ and thus phase-space trajectory is non-chaotic for given values of inter-atomic interaction. But in presence of a weak lattice potential, the LE associated with $S_x$ and $S_y$ components become positive. However, for $S_z$ component LE is always negative. Thus we say that the effective spin dynamics becomes chaotic if LE of any one or more spin components becomes positive.

{It may be interesting to note that the temporal variation of positive value of Lyapunov exponent corresponding to $S_x$ and $S_y$ grow rapidly after a certain of evolution for relatively  higher values of lattice depth and interaction strength. Thus the system becomes more chaotic.}

\section{Conclusion}
We consider spin-orbit coupled Bose-Einstein condensates  in optical lattices and study the effective spin dynamics within the framework mean-field equations. More specifically,  we take bright-bright solitons solution consisting of spin up and down particles and study both regular and nonlinear effective spin dynamics of the coupled system using variational and numerical approaches. We extract spinor part of the solution and write a nonlinear Bloch equation for the effective spin and thus find an effective magnetic field in presence of optical lattices.

 The spin angular momentum of the soliton changes by an effective magnetic field in accordance with the Bloch equation and generates a force that modifies the motion of the soliton's center-of-mass coordinate. We have demonstrated how the  interplay among the  nonlinearity,  optical lattice, and  spin-orbit coupling influence soliton's  spin expectation value and center-of-mass motion of solitons. It is seen that the optical lattice significantly affects the dynamics of spin components and the center-of-mass motion. More specifically,  the optical lattice is found to generate an extra collective oscillation with a greater frequency and amplitude in addition to the movements of solitons generated by the spin-orbit coupling.
 
 We have checked the effects of periodic variation of spin expectation value on the density profile through direct numerical simulation. It is seen that spin flipping and/ exchange occur periodically. As a result one of  the states remains highly populated during half of the period while other state remains less populated and vice versa in the next half of the period. The amplitude and frequency  of population imbalance are found to be sensitively affected by optical lattice.

{We have also examined the  effective nonlinear spin dynamics in presence of the optical lattice potential using fixed point analysis. It is seen that lattice strength can be considered as a control  parameter for changing the dynamics from regular to chaos.  The spin dynamics exhibit regular periodic oscillation in the absence of the lattice potential. However with the introduction of lattice potential the dynamics show chaotic behaviour. The is confirm from the structure of phase trajectory of spin components and the  values of Lyapunov exponents.}

\section*{Appendix}
Equations for the dynamics of different parameters are obtained from the vanishing conditions of variational derivatives of $\left\langle {\cal L}\right\rangle$ with respect to different parameters. From $\frac{\delta \left\langle {\cal L}\right\rangle}{\delta \phi_1}-\frac{\delta \left\langle {\cal L}\right\rangle}{\delta \phi_2}=0$ we get
\begin{eqnarray}
\dot{\theta}=-\frac{\alpha \pi \lambda}{a \sinh(\pi \lambda /a)} \sin \phi.
\label{eq8}
\end{eqnarray}
 Again from $\frac{\delta \left\langle {\cal L}\right\rangle}{\delta p_{1x}}+\frac{\delta \left\langle {\cal L}\right\rangle}{\delta p_{2x}}=0$ we get
\begin{eqnarray}
k_p=(k_n-\beta)\cos2\theta+\frac{d}{dt}(-x_c/a).
\end{eqnarray}
and the use of $\frac{\delta \left\langle {\cal L}\right\rangle}{\delta p_{1x}}-\frac{\delta \left\langle {\cal L}\right\rangle}{\delta p_{2x}}=0$ gives
\begin{eqnarray}
\frac{\pi  \lambda}{a} \coth \left(\frac{\pi  \lambda}{a}\right)-1=0.
\end{eqnarray} 

Similarly, from $\frac{\delta \left\langle {\cal L}\right\rangle}{\delta x_c}=0$ 
\begin{eqnarray}
\dot{k}_p=\!\frac{2\pi \alpha \lambda^2}{a \sinh(\pi \lambda/a)} \sin 2\theta \sin\phi\!+\!\frac{2\pi k_{lat}^2V_0 }{a\sinh(\pi k_{lat}/a)}\! \sin\phi
\label{eq16}
\end{eqnarray}
and from $\frac{\delta \left\langle {\cal L}\right\rangle}{\delta \theta}=0$
\begin{eqnarray}
\dot{\phi}=2k_n k_p-\frac{2\pi \alpha k_n \cos\phi}{a\sinh(\frac{\pi k_n}{a})} \cot 2\theta.
\end{eqnarray}


\begin{thebibliography}{99}
\bibitem{hall1} N. Nagaosa, J. Sinova, S. Onoda, A. H. MacDonald, and N. P. Ong, Anomalous Hall effect, Rev. Mod. Phys. \textbf{82}, 1539 (2010).
\bibitem{hall2} Di Xiao, Ming-Che Chang, and Qian Niu, Berry phase effects on electronic properties, Rev. Mod. Phys. \textbf{82}, 1959 (2010).
\bibitem{hasan} M. Z. Hasan and C. L. Kane, Colloquium: Topological insulators, Rev. Mod. Phys. \textbf{82}, 3045 (2010).
\bibitem{kane} Charles Kane and Joel Moore, Topological insulators, Physics World \textbf{24} 32 (2011).
\bibitem{scond} Xiao-Liang Qi and Shou-Cheng Zhang,Topological insulators and superconductors, Rev. Mod. Phys. \textbf{83}, 1057(2011).
\bibitem{soc1} Ming Gong, Sumanta Tewari, and Chuanwei Zhang, BCS-BEC crossover and topological phase transition in 3D spin-orbit coupled degenerate fermi gases, Phys. Rev. Lett. \textbf{107}, 195303 (2011).
\bibitem{zhan} Zhan Wu et al. Realization of two-dimensional spin-orbit coupling for Bose-Einstein condensates, Science \textbf{354}, 83(2016).
\bibitem{thomas} T. M. Bersano, J. Hou, S. Mossman, V. Gokhroo, Xi-Wang Luo, K. Sun, C. Zhang, and P. Engels, Experimental realization of a long-lived striped Bose-Einstein condensate induced by momentum-space hopping, Phys. Rev. A \textbf{99}, 051602(R) (2019).
{\bibitem{chen} Li Chen, Han Pu, and Yunbo Zhang, Spin-orbit angular momentum coupling in a spin-1 Bose-Einstein condensate, Phys. Rev. A \textbf{93}, 013629(2016).
\bibitem{balaz} I. Vasic and A. Balaz, Excitation spectra of a Bose-Einstein condensate with an angular spin-orbit coupling, Phys. Rev. A \textbf{94}, 033627 (2016).}
\bibitem{xu} Yong Xu, Yongping Zhang, and Biao Wu, Bright solitons in spin-orbit-coupled Bose-Einstein condensates, Phys. Rev. A \textbf{87}, 013614 (2013).
\bibitem{morsch} O. Morsch and M. Oberthaler, Dynamics of Bose-Einstein condensates in optical lattices, Rev. Mod. Phys. \textbf{78}, 179 (2006).
\bibitem{giovanni} Giovanni Italo Martone, Bose-Einstein condensates with Raman-induced spin-orbit coupling: An overview, Eur. Phys. Lett. \textbf{143}  25001 (2023).
\bibitem{wen} L. Wen, Q. Sun, Yu Chen, Deng-Shan Wang, J. Hu, H. Chen, W.-M. Liu, G. Juzeliunas, Boris A. Malomed, and An-Chun Ji, Motion of solitons in one-dimensional spin-orbit-coupled Bose-Einstein condensates, Phys. Rev. A \textbf{94}, 061602(R) (2016).
\bibitem{wen1} L. Wen, Xiao-Fei Zhang, Ai-Yuan Hu, J. Zhou, P. Yu, L. Xia , Q. Sun, An-Chun Ji, Dynamics of bright–bright solitons in Bose-Einstein condensate with Raman-induced one-dimensional spin–orbit coupling, Ann.  Phys. \textbf{390}, 180 (2018).
%
\bibitem{swarup} S. K. Sarkar, T. Mishra, P. Muruganandam, and P. K. Mishra, Quench-induced chaotic dynamics of Anderson-localized interacting Bose-Einstein condensates in one dimension, Phys. Rev. A \textbf{107}, 053320 (2023).
%
\bibitem{zhang} Zhang Jin-Yi et al. Collective Dipole Oscillations of a Spin-Orbit Coupled Bose-Einstein Condensate, Phys. Rev. Lett. \textbf{109}, 115301 (2012).
%
\bibitem{zhang1} S. Zhang, C. He, E. Hajiyev, Z. Ren, B. Song, G.-B. Jo, Collective dipole oscillations of a spin-orbit coupled Fermi gas, Sci. Rep. \textbf{8}, 18005 (2018).
\bibitem{chaos0} K. K. Dey and G. A. Sekh, Coupled matter-wave solitons on oscillating reflectors under the effects of gravity, Chaos \textbf{32}, 083149 (2022).
%
\bibitem{chaos1} Alan Wolf, Jack B. Swift, Harry L. Swinney, John A. Vastano, Determining Lyapunov exponents from a time series, Physica D {\bf 16}, 285 (1985).
%
\bibitem{chaos2}  K. K. Dey and  G. A. Sekh, Effects of random excitations on the dynamical response of duffing systems. J. Stat. Phys. \textbf{182}, 18 (2021).
%
\bibitem{sumaita} Sumaita Sultana and G. A. Sekh, Josephson-type oscillations in spin-orbit coupled Bose-Einstein condensates with nonlinear optical lattices, Phys. Lett. A {\bf 488}, 1291379 (2023).
%
\bibitem{bloch} Fatkhulla Kh. Abdullaev, Marijana Brtka, Arnaldo Gammal, Lauro Tomio, Solitons and Josephson-type oscillations in Bose-Einstein condensates with spin-orbit coupling and time-varying Raman frequency, Phys. Rev. A \textbf{97}, 053611 (2018).
%
\bibitem{eugen} Eugen Merzbacher, {\it Quantum Mechanics}, 3rd Edt., Wiley (1998).
%
{\bibitem{ravisankar} R. Ravisankar, D. Vudragovic, P. Muruganandam, A. Balaz, S. K. Adhikari, Spin-1 spin–orbit- and Rabi-coupled Bose–Einstein condensate solver, Comput. Phys. Commun. \textbf{259}, 107657 (2021).
%
\bibitem{muruganandam} P. Muruganandam, A. Balaz, S. K. Adhikari, OpenMP solver for rotating spin-1 spin–orbit- and Rabi-coupled Bose–Einstein condensates, Comput. Phys. Commun. \textbf{264}, 107926 (2021).}
%
\bibitem{shs} Steven H. Strogatz, Nonlinear Dynamics and Chaos: With Applications to Physics, Biology, Chemistry, and Engineering, 2nd Edition, CRC Press (2015).
%
\bibitem{rossler} O. E. R\"ossler, An equation for continuous chaos, Phys. Lett. A {\bf 57}, 397 (1976).
%
\end{thebibliography}
\end{document}